\documentclass[prc,floatfix,groupedaddress,nofootinbib,showpacs,preprintnumbers,
amsmath,amssymb,amsfonts,superscriptaddress,widetable] {revtex4-1}
\usepackage{graphicx}% Include figure files
\usepackage{dcolumn}%Align table columns on decimal point
\usepackage{mathrsfs,amsmath,amssymb}
\usepackage{bm}% bold math
\usepackage{graphicx}% Include figure files
\usepackage{dcolumn}% Align table columns on decimal point
\usepackage{bm}% bold math
\usepackage{array}
\usepackage{bigstrut}
\usepackage[usenames]{color}
\begin{document}

\title{Refining mass formulas for astrophysical applications: \\
a Bayesian neural network approach}
%Nuclear Driplines and Separation Energies Predictions \\ 
%for r-process Nucleosynthesis: Bayesian Neural Networks Method
\author{R. Utama}
\email{ru11@my.fsu.edu} 
\affiliation{Department of Physics, Florida State University, Tallahassee, FL 32306} 
%\author{Wei-Chia Chen}
%\email{wc09c@my.fsu.edu} 
%\affiliation{Department of Physics, Florida State University, Tallahassee, FL 32306}
\author{J. Piekarewicz}
\email{jpiekarewicz@fsu.edu}
\affiliation{Department of Physics, Florida State University, Tallahassee, FL 32306}
\date{\today}
\begin{abstract}
\begin{description}
 \item[Background]  Exotic nuclei, particularly those near the driplines, are at the core 
 		                of one of the fundamental  questions driving nuclear structure 
		                and astrophysics today: \emph{what are the limits of nuclear 
		                binding?} Exotic nuclei play a critical role in both informing 
		                theoretical models as well as in our understanding of the origin 
		                of the heavy elements.

\item[Purpose]        To refine existing mass models through the training of an 
	                       artificial neural network that will mitigate the large model 
	                       discrepancies far away from stability. 

\item[Methods]        The basic paradigm of our two-pronged approach is an existing 
	                       mass model that captures as much as possible of the underlying
	                       physics followed by the implementation of a Bayesian Neural 
	                       Network (BNN) refinement to account for the missing physics. 
	                       Bayesian inference is employed to determine the parameters of 
	                       the neural network so that model predictions may be accompanied 
	                       by theoretical uncertainties.

\item[Results]	      Despite the undeniable quality of the mass models adopted in this
	                       work, we observe a significant improvement (of about 40\%) after 
	                       the BNN refinement is implemented. Indeed, in the specific case of 
	                       the Duflo-Zuker mass formula, we find that the rms deviation relative 
	                       to experiment is reduced from 
	                       $\sigma_{\rm rms}\!=\!0.503$\,MeV to 
	                       $\sigma_{\rm rms}\!=\!0.286$\,MeV.	                       
	                       These newly refined mass tables are used to map the neutron drip 
	                       lines (or rather ``dripbands") and to study a few critical $r$-process 
	                       nuclei. 
	                      
\item[Conclusions]  The BNN approach is highly successful in refining the predictions of
                                existing mass models. In particular, the large discrepancy displayed 
                                by the original ``bare" models in regions where experimental data is 
                                unavailable is considerably quenched after the BNN refinement. This 
                                lends credence to our approach and has motivated us to publish refined 
                                mass tables that we trust will be helpful for future astrophysical 
                                applications.
                                      
\end{description}
\end{abstract}
%02.50.-r   %Probability theory, stochastic processes, and statistics
%02.60.Pn %Numerical optimization
%21.10.Dr  %Binding energies and masses
%21.10.Ft   %Charge distribution
%21.10.Gv %nucleon distributions
%21.60.Jz %Nuclear Density Functional Theory and extensions
%21.65.-f  %Nuclear matter
%21.65.Cd %Asymmetric matter, neutron matter
% 21.65.Mn % Equations of state of nuclear matter
%21.65.Ef %Symmetry energy
%24.10.Jv %Relativistic models
%24.30.Cz %Giant resonances
%24.80.+y %nuclear tests of fundamental interactions and symmetries
%25.30.Bf %Elastic electron scattering
%26.50.+x %Nuclear physics aspects of novae, supernovae, and other explosive environments
%26.60.Gj %Neutron-star crust
%26.60.Dd %Neutron-star core
%26.60.Kp %Neutron-star EOS
%26.60.-c  %Nuclear matter aspects of Neutron stars
%97.60.Jd %Neutron stars

\maketitle

%%%%%%%%%%%%%%%%%%%%%%%%%%%%%%%%
\section{Introduction}
\label{intro}

Where do the chemical elements come from and how did they 
evolve is one of the central questions animating nuclear science
today\,\cite{LongRangePlan}. Stars heavier than about eight solar
masses ($M_{\star}\!\gtrsim\!8\,M_{\odot}$) reach high enough core
temperatures to support the formation of ever-heavier chemical
elements by thermonuclear fusion: from helium all the way to
iron. Yet, once the iron-peak elements are synthesized, thermonuclear
burning stops abruptly with the formation of an inert iron core that
will collapse once its mass exceeds the Chandrasekhar
limit\,\cite{Clayton:1983,Phillips1998,Iliadis:2007}. This situation
naturally begs the question: how did the elements heavier than iron form? 
Whereas the slow neutron capture process (``$s$-process'') in
asymptotic giant branch stars is believed to be responsible for the
formation of about half of the heavy elements beyond iron (such as
strontium, barium, and lead) identifying the precise site (or sites) of
the rapid neutron capture process (``$r$-process'') responsible for
the remaining half of of the heavy elements (such as gold, platinum, 
and uranium) remains elusive\,\cite{Burbidge:1957vc,Clayton:1983,
Wallerstein:1997}. Indeed, \emph{``How were the elements from iron 
to uranium made?"} has been identified as one of the eleven science 
questions for the new century\,\cite{QuarksCosmos:2003}.

Understanding $r$-process nucleosynthesis is a fascinating and
challenging multidisciplinary problem.  Progress in this area demands
detailed knowledge of a host of nuclear-structure observables---often
at the limits of nuclear existence---such as masses, neutron capture
rates, and nuclear beta
decays\,\cite{Burbidge:1957vc,Petermann:2010,Mumpower:2015ova}. 
Given that the $r$-process develops under extreme astrophysical 
conditions such as those found in the merger of two neutron stars, neutron
capture on seed nuclei occurs on a time scale that is much faster than
the competing beta-decay rates. This drives the $r$-process far away
from stability where little is known about the thousands of exotic
nuclear species that participate in these reactions; see 
Refs.\,\cite{Wallerstein:1997, Petermann:2010,Mumpower:2015ova} 
and references contained therein. In an effort to mitigate this problem,
Mumpower and collaborators have performed sensitivity studies to
identify the nuclear inputs ({\sl e.g.,} nuclear masses) that have the
greatest impact on the
$r$-process\,\cite{Mumpower:2015ova,Mumpower:2015zha,Mumpower:2015hva}. 
Whereas these studies suggest that some of the ``most influential'' nuclear
masses are within reach of future radioactive beam facilities, it is also
recognized that some others will likely remain beyond experimental
reach. Hence, theoretical guidance becomes absolutely
critical. Unfortunately, theoretical mass models that agree in the
vicinity of measured nuclear masses, disagree strongly---often by
several MeV---far away from stability. This is particularly
troublesome given that some sensitivity studies suggest that resolving
the abundance pattern will require reducing mass-model uncertainties
to $\lesssim\!100$\,keV\,\cite{Mumpower:2015hva}.

The emergence of the $r$-process pattern is highly sensitive to
nuclear masses in the vicinity of neutron magic numbers $N\!=\!50,
82$, and $126$. Interestingly, nuclear masses around closed shells
$N\!=\!50$ and $82$ also have a profound impact on the composition of
a different astrophysical system: the outer crust of a neutron
star. Given that at the low densities of relevance to the outer crust
($10^{4}$\,--\,$10^{11}\,{\rm g/cm^{3}}$) the average inter-nucleon
separation is considerably larger than the range of the nuclear
interaction, it becomes energetically favorable for nucleons to
cluster into individual nuclei that arrange themselves in a
crystalline lattice that is immersed in a uniform free
Fermi gas of electrons\,\cite{Baym:1971pw}. And whereas the underlying
dynamics is relatively simple, the exotic composition of the outer
stellar crust is also highly sensitive to nuclear masses in regions
where experimental information is not yet
available\,\cite{RocaMaza:2008ja}.  Thus, as in the case of the
$r$-process, understanding the composition of the stellar crust relies
heavily on theoretical extrapolations into unknown regions of the
nuclear chart.

In an effort to lessen the impact of such inevitable extrapolations
we have recently intoduced a Bayesian Neural Network (BNN) 
approach for the calculation of nuclear
masses\,\cite{Utama:2015hva,Piekarewicz:2016akv} and  
charge radii\,\cite{Utama:2016rad}. The novel framework proposed in
Ref.\,\cite{Utama:2015hva} consists of a combined scheme that relies
on accurate theoretical predictions that are subsequently refined by
training a suitable artificial neural network on the \emph{residuals} between 
the experimental data and the theoretical predictions. In essence, 
the central tenet of our approach is to include as much physics as 
possible in the underlying nuclear model and then rely on a BNN 
refinement to recover most of the physics that is missing from the 
model. In our earlier work we were able to identify several virtues 
of such combined approach. First, by using a randomly selected set
of experimentally known masses to train the neural network, we 
observed a significant improvement in the predictions of the 
remaining known masses that were not included in the training 
set---even for some of the most sophisticated mass models available 
in the literature. Second, it is well known that theoretical mass models 
of similar quality agree 
in regions where masses are known, but differ widely (by as much 
as tens of MeVs) in regions where experimental data is not yet 
available\,\cite{Blaum:2006}. However, after implementing the BNN 
refinement we found that the large differences in the model 
predictions were significantly reduced. Finally, given  the ``Bayesian" 
character of the approach, the refined predictions were 
accompanied by properly estimated theoretical 
errors\,\cite{Utama:2015hva}.

In our earlier work we have successfully tested the BNN paradigm in the
context of nuclear masses of relevance to the outer crust of neutron 
stars\,\cite{Utama:2015hva,Piekarewicz:2016akv}. It is the main goal 
of the present paper to provide updated mass tables that encompass the 
entire nuclear chart. As several sophisticated and successful mass 
tables already exist\,\cite{Moller:1981zz,Moller:1988,Moller:1993ed,
Duflo:1995,Goriely:2010bm,Kortelainen:2010hv,Erler:2012qd}, the
initial phase of our program---namely, the selection of an underlying 
model that incorporates as much physics as possible---is essentially 
complete. Thus, the remaining task is to implement the BNN refinement 
on these highly successful models. Our hope is that the BNN refinement 
will improve the original mass models by reducing the large systematic 
uncertainties and by providing realistic statistical errors. Together with the 
publication of these updated mass tables, we will also make available
(with associated theoretical uncertainties) related observables 
that may be more suitable for certain astrophysical applications, such 
as one- and two-nucleon separation energies. It is our hope that these 
tables will contribute to further our understanding of astrophysical 
phenomena as well as to constrain theoretical models of nuclear structure. 
Ultimately, of course, any progress in theory is strongly coupled to 
experimental advances. Indeed, we are at the dawn of a new era where 
rare isotope facilities will probe the limits of nuclear existence and in so 
doing will provide critical guidance to theoretical models. And although 
some of the required theoretical extrapolations will take us far into 
regions of the nuclear chart that are unlikely to be explored even at the 
most sophisticated facilities, measurements of even a few exotic short-lived 
isotopes are of critical importance for the improvement of theoretical
models. 

The manuscript has been organized as follows. In the next section we
provide some additional details on the strong synergy between nuclear
structure and astrophysics. As the BNN approach has been presented 
elsewhere\,\cite{Utama:2015hva}, we limit the discussion on the 
formalism to a brief outline of the method and its implementation in 
Sec.\,\ref{Formalism}. Next, in Sec.\,\ref{Results} we focus on the 
improvement to several existing mass models after the BNN refinement. 
Only in the case of the 10-parameter version of the Duflo-Zuker mass 
model\,\cite{Duflo:1995} we recalibrate the parameters in order to extract 
the associated covariance matrix that encodes statistical uncertainties 
and correlations among the model parameters. In this way the overall 
theoretical error will have its origin in two sources: (a) the uncertainty 
in calibration of the ``bare'' model parameters and (b) the errors emerging 
from the BNN refinement. Also presented in this section are results for 
proton and neutron drip lines as well as a few masses of 
particular relevance to the $r$-process. 
Finally, we conclude in Sec.\,\ref{Conclusions} with a summary of 
our important findings.

%%%%%%%%%%%%%%%%%%%%%%%%%%%%%%%%
%%%%%%%%%%%%%%%%%%%%%%%%%%%%%%%%
\section{Astrophysical Motivation}
\label{AstroMot}

Although of fundamental and high intrinsic nuclear-physics value, modern 
nuclear mass tables find today their best expression in astrophysical 
applications. In particular, nuclear masses are of paramount importance in 
understanding nucleosynthesis in hot stellar environments and the crustal 
composition of cold neutron stars. In what follows we provide a brief 
description of these two scenarios underscoring the critical role of nuclear 
masses far away from equilibrium.

\subsection{Neutron capture and photo-dissociation: $(\gamma,n)$ equilibrium}

Although modern $r$-process simulations make no assumptions on whether 
stellar conditions---such as temperature, density, and neutron fraction---are 
favorable to maintain the reaction $(n,\gamma)\!\rightleftharpoons\!(\gamma,n)$ 
in thermodynamic equilibrium, the outcome of such network calculations suggests 
that under many astrophysical scenarios equilibrium is indeed attained, at least 
during the early stages. Under such an astrophysical scenario, the final abundance 
pattern follows solely from statistical equilibrium and nuclear physics. In particular, 
the pattern along a given isotopic chain is set by the temperature ($T$), the neutron 
density ($N_{n}$), and the one-neutron separation energy ($S_{n}$). In essence, 
once thermodynamic equilibrium has been established in a given astrophysical 
environment, the final abundance pattern along an isotopic chain is entirely determined 
by nuclear masses, both near and far from the valley of stability. The ratio of yields of 
neighboring isotopes at equilibrium is set by the equality of the chemical potential of 
the competing species. Given that $\mu_{\gamma}\!\equiv\!0$ one obtains,
%%%
\begin{equation}
 \mu(Z,A) + \mu_{n} =  \mu(Z,A+1).
\label{ChemEquil}
\end{equation} 
%%%
In the limit of low neutron density and high temperature, so that each component may 
be treated as a classical ideal gas, the condition of chemical equilibrium is encoded 
in the Saha equation\,\cite{Clayton:1983,Phillips1998,Iliadis:2007}:
%%%
\begin{equation}
 \frac{Y(Z,A+1)}{Y(Z,A)}  = \frac{G(Z,A+1)}{2G(Z,A)}N_{n}\lambda_{n}^{3}(T)
 \exp\!\left( \frac{S_{n}(Z,A+1)}{k_B T}\right).
\label{NucSaha}  
\end{equation} 
%%% 
Here $Y(Z,A)$ and $G(Z,A)$ denote the isotopic abundance and partition function of 
the seed nucleus, and $\lambda_{n}(T)\!=\!\sqrt{2\pi/m_{n}k_{\rm B}T}$ is the de~Broglie 
thermal wavelength of the neutron. Often $N_{Q}\!=\!\lambda^{-3}_{n}(T)$ is referred 
to as the quantum concentration; if $N_{n}\!\ll\!N_{Q}$ then the system behaves 
classically. The nuclear dynamics is imprinted in the neutron separation energy
%%%
\begin{equation}
 S_{n}(Z,A+1)\!\equiv\!M(Z,A) + m_{n} - M(Z,A+1)  = B(Z,A+1) - B(Z,A), 
\label{SepEne}  
\end{equation} 
%%% 
with $B(Z,A)$ the total nuclear binding energy. Note that the relative abundance is
\emph{exponentially} sensitive to errors in the neutron separation energy. For example,
at a canonical stellar temperature of $T_{9}\!=\!10^{9}\,{\rm K}$, a relatively ``modest'' 
error  in the separation energy of $0.1$\,MeV translates into an error in the relative 
abundance of about a factor of three. Eventually, chemical equilibrium is lost and the 
final abundance pattern is dictated by a series of $\beta$ decays back to stability. 
Here too nuclear masses are of critical importance since the phase-space factor for 
beta decay is determined by the reaction $Q$-value:
$Q_{\beta}\!=\!M(Z,A)\!-\!M(Z+1,A)$.

\subsection{Crustal composition of a neutron star}

Another powerful connection between astrophysics and nuclear physics that is highly
sensitive to nuclear masses involves the crustal composition of a neutron star, particularly 
its outer crust\,\cite{Baym:1971pw,Haensel:1989,Haensel:1993zw,Ruester:2005fm,
RocaMaza:2008ja,RocaMaza:2011pk}. The crust is interesting because the dynamics is 
simple yet subtle. Simple, because nuclear masses is the only ingredient driving the
composition of the outer crust. Subtle, since the crustal composition emerges from a 
delicate dynamics between the electronic energy and the nuclear symmetry energy.
Indeed, at the densities of relevance to the outer crust, it is energetically favorable for 
nucleons to cluster into nuclei that arrange themselves in a body-centered cubic lattice 
that itself is immersed in a neutralizing electron background\,\cite{Baym:1971pw}. At zero 
temperature and fixed pressure, the dynamics of the outer crust is encoded in the following 
expression for the chemical potential (or Gibbs free energy per nucleon) of the system:
%%%
 \begin{equation}
   \mu(Z,A; P) = \frac{M(Z,A)}{A} + \frac{Z}{A}\mu_{e} -
   \frac{4}{3}C_{l}\frac{Z^2}{A^{4/3}}\,p_{{}_{\rm F}}. 
   \label{ChemPot}
\end{equation}
%%%
The first term---which is independent of the pressure---depends exclusively on the mass per 
nucleon of the ``optimal'' nucleus populating the crystal lattice. The second  term $\mu_{e}$ 
is the chemical potential of a relativistic Fermi gas of electrons. Finally, the last  term provides 
the relatively modest, although by no means negligible, lattice contribution 
($C_{l}\!=\!3.40665\!\times\!10^{-3}$) to the chemical potential. Note that both the electronic 
%%%
 \begin{equation}
  \mu_{e}\!=\!\sqrt{
  \left(\frac{Z}{A}\right)^{\!\!2/3}\hspace{-5pt}p_{\rm F}^{2}+m_{e}^{2}}\,,
  \label{eChemPot}
\end{equation}
%%%
and lattice contributions have been written in terms of the Fermi momentum 
$p_{{}_{\rm F}}= (3\pi^2n)^{1/3}$ (or equivalently the baryon density $n$) rather than
the pressure. The connection between the baryon density and the pressure is obtained
through the equation of state. That is\,\cite{RocaMaza:2008ja},
%%%
 \begin{equation}
   P(Z,A; n) =
   \frac{m_{e}^{4}}{3\pi^{2}} 
   \left(x_{{}_{\rm F}}^{3}y_{{}_{\rm F}} - 
   \frac{3}{8}\left[x_{{}_{\rm F}}y_{{}_{\rm F}}
   \Big(x_{{}_{\rm F}}^{2}+y_{{}_{\rm F}}^{2}\Big)
  -\ln(x_{{}_{\rm F}}+y_{{}_{\rm F}}) \right]\right)
  -\frac{n}{3}C_{l}\frac{Z^2}{A^{4/3}}\,p_{{}_{\rm F}},
 \label{Pressure}   
\end{equation}
%%%
where 
%%%
 \begin{equation}
   x_{{}_{\rm F}}\!=\!\frac{p^{(e)}_{{}_{\rm F}}}{m_{e}} =
   \left(\frac{Z}{A}\right)^{1/3}\!\frac{p_{{}_{\rm F}}}{m_{e}} 
   \quad{\rm and}\quad
   y_{{}_{\rm F}}\!=\!\sqrt{1\!+\!x_{{}_{\rm F}}^{2}}\,, 
 \label{ScaledxF} 
\end{equation}
%%%   
are the scaled electronic Fermi momentum and Fermi energy, respectively. Given that 
the outer crust spans nearly seven orders of magnitude in density, from about 
$10^{4} {\rm g/cm^{3}}$ to $10^{11} {\rm g/cm^{3}}$, changes in the nuclear composition
with density (or pressure) are interesting despite the simplicity of the underlying dynamics. 
For example, at the top of the outer crust where the pressure is low and so is the density, 
the electronic contribution to the chemical potential is negligible, so it is favorable to populate 
the crystal lattice with the nucleus having the the lowest mass per nucleon in the entire 
nuclear chart: ${}^{56}$Fe. However, as the pressure and the density increase, it 
becomes energetically advantageous for the system to lower its electron fraction 
$Y_{e}\!=\!Z/A$ via electron capture on the protons. As a consequence, ${}^{56}$Fe 
ceases to be the optimal nucleus due to the presence of a uniform sea of neutralizing 
electrons whose chemical potential increases rapidly with density. Thus, the essential 
physics of  the outer crust involves a competition between an electronic contribution 
that favors $Y_{e}\!=\!0$ and the nuclear symmetry energy that instead favors 
$Y_{e}\!\simeq\!1/2$.

Ultimately, computing the nuclear composition of the stellar crust requires precise 
knowledge of nuclear masses over three well-defined regions of the nuclear chart.
At the top of the crust the electronic contribution to the chemical potential is small
to moderate, so the isotopes of relevance are located around the stable iron-nickel 
region where nuclear masses are very well known. As the proton fraction 
becomes too low and the symmetry energy large, it becomes energetically favorable 
for the system to jump into the $N\!=\!50$ region. The nuclei of relevance in this region 
lie at  the boundary between those whose masses are accurately known ({\sl e.g.,} 
${}^{90}$Zr, ${}^{88}$Sr, and ${}^{86}$Kr) and those that are poorly known (such as 
${}^{78}$Ni). We should mention that until very recently the mass of ${}^{82}$Zn was 
not known. Yet, the mass of ${}^{82}$Zn was determined fairly recently at the 
ISOLDE-CERN  facility, leading to an interesting modification of the crustal 
composition\,\cite{Wolf:2013ge}. Finally, the third region comprising the bottom 
layers of the outer crust requires knowledge of nuclear masses at the $N\!=\!82$ 
shell closure. Depending on the particular mass model, the nuclei of relevance 
span the region from ${}^{132}$Sn  ($Z\!=\!50$) all the way down to ${}^{118}$Kr 
($Z\!=\!36$)\,\cite{RocaMaza:2011pk}. For this region theoretical extrapolations 
are unavoidable as little or no experimental information is 
available\,\cite{Utama:2015hva}. Indeed, ${}^{118}$Kr is 21 neutrons away from 
the last measured isotope with a well measured mass\,\cite{AME:2012}.

%%%%%%%%%%%%%%%%%%%%%%%%%%%%%%%%
%%%%%%%%%%%%%%%%%%%%%%%%%%%%%%%%
\section{Formalism}
\label{Formalism}

\subsection{Bayesian neural networks}

The novel theoretical approach that we advocate here aims to refine some 
existing mass models through a BNN approach\,\cite{Neal1996}. Given the
proven success of modern mass models, the BNN refinement is implemented 
by training a suitable neural network on the \emph{residuals} between the 
experimental data and the ``bare'' ({\sl i.e.,} before refinement) theoretical 
predictions. Our ultimate goal is to generate a \emph{universal 
approximator}\,\cite{Cybenko:1989,Hornik:1989}; that is, a neural network 
that can provide an ``educated'' extrapolation into unexplored regions of the 
nuclear chart with properly quantified theoretical uncertainties. A detailed 
description of the origins and development of Bayesian neural networks 
goes beyond the scope of this paper; for a detailed exposition see 
Refs.\,\cite{Neal1996,Bishop1995,Haykin1999,Vapnik1998}. Thus, as we 
have done elsewhere\,\cite{Utama:2015hva,Piekarewicz:2016akv,Utama:2016rad}, 
we limit ourselves to highlight the main features of the approach. Before we 
do so, however, we note that the idea of using artificial neural networks in 
nuclear physics---mainly to estimate unknown properties of exotic nuclei of 
relevance to astrophysics---started in the early 90s with the work of Clark 
and collaborators\,\cite{Gazula:1992,Gernoth:1993,Gernoth:1995,Clark:1999} 
and continues up to this day\,\cite{Athanassopoulos:2003qe,
Athanassopoulos:2005rc,Clark:2006ua,Costiris:2009,Bayram:2013hi} with 
more sophisticated applications. 

Statistical inference 
based on Bayes' theorem---as applied in this work---connects two critical pieces of 
information: (a) a \emph{prior} hypothesis reflecting beliefs that one has acquired 
through experience or previous empirical information and (b) an \emph{improvement} 
to the prior hypothesis by both adopting and adapting new evidence ({\sl e.g.,} 
experimental data). In this context Bayes' theorem may be written as\,\cite{Stone:2013}:
%%%
\begin{equation}
 p(\omega|x,t)=\frac{p(x,t|\omega)p(\omega)}{p(x,t)},
 \label{BayesRule}
\end{equation} 
%%%
where $p(\omega)$ is the prior distribution of the model parameters $\omega$ and
$p(x,t|\omega)$ is the ``likelihood" that a given model $\omega$ describes the new 
evidence $t(x)$. The product of the prior and the likelihood form the \emph{posterior} 
distribution $p(\omega|x,t)$ that encodes the probability that a given model describes 
the data $t(x)$. In essence, the posterior represents the improvement to $p(\omega)$
as a result of the new evidence $p(x,t|\omega)$. Note that the ``marginal likelihood" 
$p(x,t)$ is independent of the model parameters $\omega$, so for our purposes it 
may be regarded as an overall normalization factor. To define the likelihood we start
by introducing an objective (or cost) function in terms of a least-squares fit to the 
empirical data. That is,
%%%
\begin{equation}
 \chi^2(\omega)=\sum_{i=1}^{N}
 \left(\frac{t_i-f(x_i,\omega)}{\Delta t_{i}}\right)^2,
 \label{Chi2}
\end{equation}    
%%%
where $N$ is the total number of data points, $t_{i}\!\equiv\!t(x_{i})$ is the empirical
value of the target evaluated at the $i$th input $x_{i}$, $\Delta t_{i}$ is the associated 
error, and the universal approximator $f(x,\omega)$ depends on both the input data 
and the model parameters $\omega$; see Eq.(\ref{ANN}) below. From such an
objective function the likelihood is customarily defined as 
%%%
\begin{equation}
 p(x,t|\omega)=\exp\big(\!-\!\chi^{2}(\omega)/2\big).
 \label{likelihood}
\end{equation} 
%%%
In the particular case of interest here, {\sl i.e.,} nuclear masses, the input $x\!\equiv\!(Z,A)$ 
represents the charge and mass number of the nucleus and $t(x)\!\equiv\!\delta M(Z,A)$ the 
mass residual between the experimental data and the theoretical predictions. Note that
maximizing the likelihood $p(x,t|\omega)$ provides the \emph{maximum likelihood estimation} 
of the model parameters. 

\begin{figure}[ht]
\vspace{-0.05in}
\includegraphics[width=0.45\columnwidth] {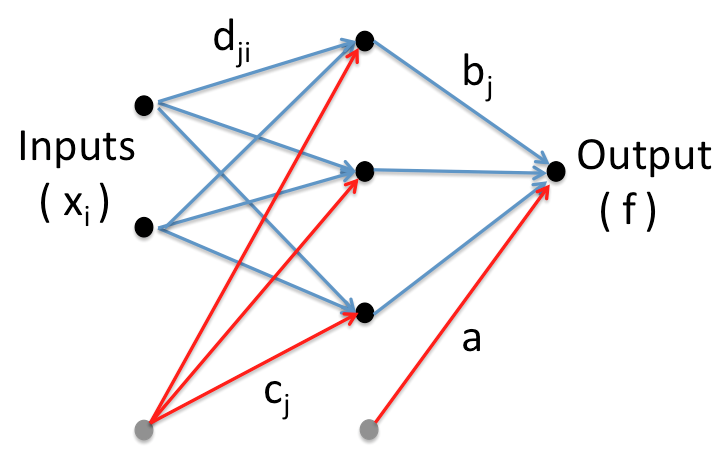}
\caption{An example of a feed-forward neural network with a single 
hidden layer consisting of three nodes. In our case, the two inputs
that define the nucleus of interest are $Z$ and $A$, and a single 
output provides an estimate of $\delta M(Z,A)$, namely, the 
discrepancy between the bare theoretical prediction and the 
experimental value.}
\label{Fig1}
\end{figure}

The neural network function $f(x,\omega)$ adopted here has the following ``sigmoid" form:
%%%
\begin{equation}
  f(x,\omega)=a+\sum_{j=1}^H b_j \tanh\left(c_j+\sum_{i=1}^I d_{ji} x_i\right),
  \label{ANN}
\end{equation}
%%%  
where the model parameters (or ``connection weights'') are collectively given by 
$\omega\!=\!\big\{a,b_{j},c_{j},d_{ji}\big\}$, $H$ is the number of hidden nodes, and $I$ 
is the number of inputs. The ``universal approximation theorem'' states  that such a 
neural network can accurately represent a wide variety of functions; $\tanh$ is a 
common form of the sigmoid activation function that controls the firing of the artificial 
neurons\,\cite{Cybenko:1989,Hornik:1989}. See Fig.\,\ref{Fig1} for a simple depiction 
of a feed-forward neural network consisting of a single hidden layer with three nodes.

%%%%%%%%%%%%%%%%  Figure 1  %%%%%%%%%%%%%%%
%%%%%%%%%%%%%%%%  Figure 1  %%%%%%%%%%%%%%%

Given the complexity of the posterior distribution $p(\omega|x,t)$, we adopt  
Markov Chain Monte Carlo (MCMC) sampling to generate a faithful equilibrium 
distribution. Once a significant number of samples has been generated, reliable 
estimates for both the average and variance of $\delta M(Z,A)$ are obtained. 
That is, 
%%%
\begin{subequations}
\begin{alignat}{4}
  \langle f_{n}\rangle &=\frac{1}{K}\sum_{k=1}^{K} f(x_{n},\omega _{k})
  \label{Avgfn}, \\
  \langle f^{2}_{n}\rangle &=\frac{1}{K}\sum_{k=1}^{K} f^{2}(x_{n},\omega _{k})
  \label{Avgfn2}, \\
 \Delta f_{n} &= \sqrt{\langle f_{n}^{2}\rangle - \langle f_{n}\rangle^{2}},
 \label{Errorfn}
\end{alignat}
\label{MCMC}
\end{subequations}
%%%
where $x_{n}\!=\!(Z_{n},A_{n})$ represents a particular nucleus with charge $Z_{n}$ 
and mass number $A_{n}$, $K$ is the total number of Monte Carlo configurations, 
and $f(x_{n},\omega_{k})$ is the neural network estimate of $\delta M(Z_{n},A_{n})$ 
as predicted by the $k$th Monte Carlo configuration.

Finally, we conclude this section by briefly addressing the choice of prior $p(\omega)$ 
assumed in this work. Prior probabilities encode our beliefs concerning the model 
parameters and are an essential ingredient of the Bayesian paradigm. Normally, 
the prior is highly informative as it is based on our own physics biases and intuition, 
which are often well informed by prior experimental data. Unfortunately, 
whereas physics principles guide the construction of modern nuclear mass models, 
physics intuition is of no help in designing the connection weights $\omega$. 
Thus, we are forced to rely on assumptions that have been proven effective and 
reliable through mostly trial and error\,\cite{Neal1996}. Following our earlier 
work\,\cite{Utama:2015hva}, we assume all connection weights to be independent 
and adopt a Gaussian prior centered around zero and with a width determined 
as in Ref.\,\cite{Neal1996}. For an extensive discussion on the determination of 
the  ``hyperparameters'' controlling the width of the Gaussian prior see 
Refs.\,\cite{Mackay:1995,Mackay:1999}.

%%%%%%%%%%%%%%%%%%%%%%%%%%%%%%%%%%%%%%%%%%%%%%%

\section{Results}
\label{Results}

The aim of this section is to discuss the improvement to three successful mass 
models as a result of the BNN refinement. The three models under consideration 
are: (i) the 10-parameter Duflo-Zuker model (DZ10)\,\cite{MendozaTemis:2009ia}, 
(ii) the 28-parameter Duflo-Zuker model (DZ)\,\cite{Duflo:1995}, and (iii) the 
microscopic Hartree-Fock-Bogoliubov model (HF19) of  Ref.\,\cite{Goriely:2010bm}. 
In all three cases the predictions after refinement have the distinct advantage of 
being accompanied by theoretical uncertainties. However, for the simpler DZ10 
model we found instructive to re-calibrated the model-parameters, as this process 
generates a suitable covariance matrix from where statistical uncertainties and 
correlation coefficients may be computed. To reiterate, the basic paradigm of 
our two-pronged approach is to start with a robust underlying mass model that 
captures as much physics as possible followed by a BNN refinement  that will 
hopefully account for the missing physics\,\cite{Utama:2015hva}.

\subsection{The 10-parameter Duflo-Zuker Model}

The original Duflo-Zuker model containing a total of 28 parameters 
has stood the test of time\,\cite{Duflo:1995}. The DZ model, fitted to the 
set of existing nuclear masses appearing in the 1995 compilation by Audi 
and collaborators\,\cite{Audi:1995}, was enormously successful in predicting 
the more than 300 additional masses that appeared in the later AME03 
compilation\,\cite{Audi:2002rp}. Indeed, the success of the DZ model in 
accurately reproducing the masses of the more than 3,000 nuclei 
presently known is truly remarkable\,\cite{Wang:2012}. However, despite 
its undeniable success, the underlying physics of the model remains 
puzzling and difficult to unravel. The simpler 10-parameter Duflo-Zuker 
model was conceived with the sole purpose of illuminating the physics. A 
detailed study of DZ10 that aims to understand and possibly to also
improve the model has been carried out recently by Mendoza-Temis, 
Hirsch, and Zuker\,\cite{MendozaTemis:2009ia}. Even more recently, 
Doboszewski and Szpak have refitted DZ10 with the goal of quantifying 
the model uncertainties and the correlations among the model 
parameters\,\cite{Doboszewski:2014}. 

\begin{figure}[h]
\vspace{-0.05in}
\includegraphics[width=0.5\columnwidth,angle=0]{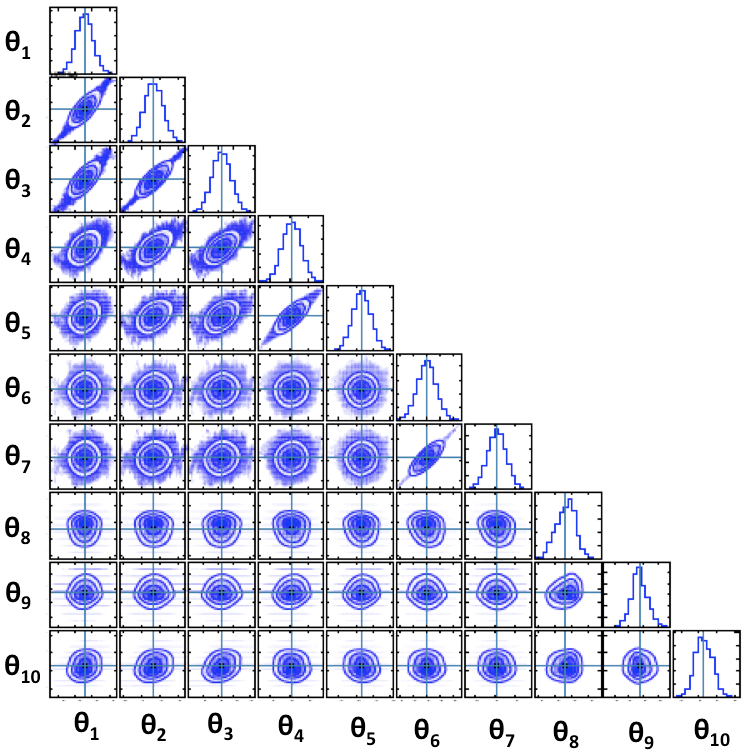}
\caption{Correlations coefficients and marginalized probability densities
(shown along the diagonal) for the 10-parameter Duflo-Zuker
model\,\cite{Duflo:1995,MendozaTemis:2009ia,Doboszewski:2014}.}
\label{Fig2}
\end{figure}

We also proceed here by recalibrating the model parameters following 
closely Ref.\,\cite{Doboszewski:2014}. To do so we start by constructing 
a likelihood function defined in terms of  the square differences between 
the DZ10 predictions and the experimental binding energies provided by 
the AME2012 compilation\,\cite{AME:2012}; we limit ourselves to the 
${}^{40}$Ca to ${}^{240}$U region.
For the initial distribution of model parameters we use an uninformative 
prior. In this way we generate a posterior probability distribution that is 
generated through Markov-Chain Monte Carlo (MCMC) sampling. 
Simulating such a posterior distribution is computationally inexpensive 
so we could afford generating one million Monte Carlo configurations,
with the first 10,000 steps used for thermalization. To avoid correlations 
among subsequent configurations, we found that an autocorrelation ``time" 
of about 20 configurations was sufficient. Overall, we used nearly 50,000 
configurations to generate the correlation plot depicted in Fig.\,\ref{Fig2} 
together with the average values and uncertainties listed in Table\,\ref{Table1}. 
For comparison, also included in Table\,\ref{Table1} are the results reported 
in Refs.\,\cite{MendozaTemis:2009ia,Doboszewski:2014}.

%%%%%%%%%%%%%%%%  Figure 2  %%%%%%%%%%%%%%%
%%%%%%%%%%%%%%%%  Figure 2  %%%%%%%%%%%%%%%

%%%%%%%%%%%%%%%%% Table I %%%%%%%%%%%%%%%%%%%%%%%%%%%
\begin{center}
\begin{table}[h]

\begin{tabular}{|c|r|r|r|}
 \hline
 Parameter & Ref\,\cite{MendozaTemis:2009ia} & Ref.\,\cite{Doboszewski:2014}\hspace{6pt}
 & This work\hspace{4pt} \\
  \hline
  $\theta_{1}\!=\!a_{3}$    &  0.707 & 0.70506(37)  & 0.70452(27)  \\
  $\theta_{2}\!=\!a_{1}$    & 17.766 & 17.749 (70)   & 17.7466(49)  \\
  $\theta_{3}\!=\!a_{2}$    & 16.314 & 16.267(2)      &16.277(17)     \\  
  $\theta_{4}\!=\!a_{4}$    & 37.515 & 37.465(4)      & 37.608(27)    \\ 
  $\theta_{5}\!=\!a_{5}$    & 53.351 & 53.23(19)      & 54.08(11)      \\ 
  $\theta_{6}\!=\!a_{7}$    & 0.478 & 0.4631(58)      & 0.4601(63)    \\ 
  $\theta_{7}\!=\!a_{8}$    & 2.183 & 2.101(30)        & 2.088(33)      \\ 
  $\theta_{8}\!=\!a_{9}$    & 0.022 & 0.02144(17)    & 0.02106(19)  \\ 
  $\theta_{9}\!=\!a_{10}$  & 41.338 & 41.5310(2)    & 41.50(22)      \\ 
  $\theta_{10}\!=\!a_{6}$  & 6.199 & 6.238(89)        & 6.455(81)      \\ 
 \hline\rule{0pt}{2.5ex}
 $\sigma$\,(MeV) & 0.554 & 0.571\hspace{12pt} & 0.552 \hspace{12pt}  \\
 \hline
\end{tabular}
\caption{Average values and uncertainties for the 10-parameter Duflo-Zuker 
model\,\cite{Duflo:1995}. Comparisons are made against similar results 
presented in Ref.\,\cite{MendozaTemis:2009ia} (without uncertainties) and 
Ref.\,\cite{Doboszewski:2014}. Here $\sigma$ represents the root-mean-square 
deviation of the predictions relative to the AME2012 compilation\,\cite{Wang:2012}.}
\label{Table1}
\end{table}
\end{center}
%%%%%%%%%%%%%%%%%%%%%%%%%%%%%%%%%%%%%%%%%%%%%%%%%%%%%%%%%%%%%%%%%

Once the theoretical predictions from the DZ10 model---or indeed any other mass 
model---have been generated, the BNN refinement proceeds by separating the 
available experimental data into two disjoint sets: a learning and a validation set. 
Again, to avoid regions of the nuclear landscape where the masses fluctuate 
too rapidly (as in the case of the lightest nuclei) we limit the experimental data set 
to the 2,000-plus well measured nuclei between $^{40}$Ca and $^{240}$U.
The learning set consists of a randomly-selected subset of nuclei contained within 
the experimental database that will be used to train the neural network, namely, to 
calibrate the connection weights as defined in Eq.(\ref{ANN}). On the other hand, 
the validation set comprises the remaining nuclei that, while still in the existent 
experimental database, were not used in the training of the network. Thus, the 
validation set provides the testbed for assessing the quality of the artificial neural 
network. If the test is successful, then one re-calibrates the network parameters 
using the entire experimental database in order to predict the masses of nuclei that 
have not yet been measured, yet are essential for astrophysical applications. 
Given that the limits of nuclear binding is one of the key science driver animating 
nuclear science today\,\cite{Erler:2012}, besides generating refined mass tables 
we also provide tables for one- and two-nucleon separation energies:
%%%
\begin{subequations}
\begin{eqnarray}
S_{n}(Z,N)   & \equiv & M(Z,N-1) + m_{n} - M(Z,N) = B(Z,N) - B(Z,N-1), 
   \label{Sn} \\ 
S_{p}(Z,N)   & \equiv & M(Z-1,N) + m_{p} - M(Z,N) = B(Z,N) - B(Z-1,N), 
   \label{Sp} \\ 
S_{2n}(Z,N) & \equiv & M(Z,N-2) + 2m_{n} - M(Z,N) = B(Z,N) - B(Z,N-2), 
   \label{S2n} \\  
S_{2p}(Z,N) & \equiv & M(Z-2,N) + 2m_{p} - M(Z,N) = B(Z,N) - B(Z-2,N).
   \label{S2p} \\ 
 \label{Drip}
\end{eqnarray} 
 \end{subequations} 
%%%
All these BNN-improved tables are generated from the same ensemble of Monte 
Carlo configurations, so that theoretical uncertainties may be reliably attached to 
each individual prediction.

%%%%%%%%%%%%%%%%  Figure 3  %%%%%%%%%%%%%%%
\begin{figure}[h]
\vspace{-0.05in}
\includegraphics[width=0.85\columnwidth]{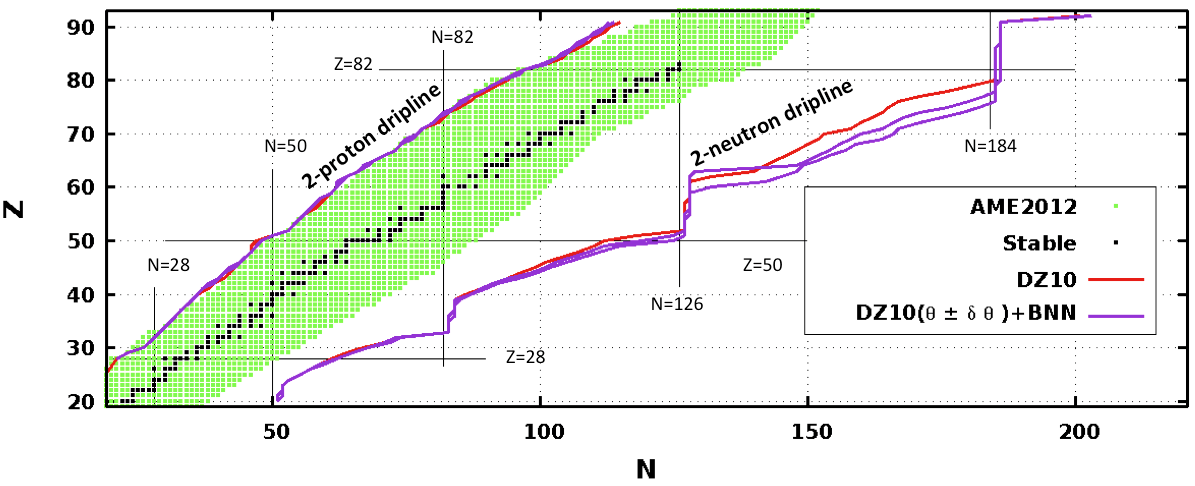}
\caption{Proton and neutron drip lines as predicted by both the bare and
BNN-refined 10-parameter Duflo-Zuker (DZ10) model. In the case of the 
BNN predictions, the neutron drip line evolves into a ``drip band'' due to 
statistical uncertainties. Also shown (with green points) is the AME2012 
data set\,\cite{AME:2012} with the stable nuclei displayed with black 
points.}
\label{Fig3}
\end{figure}
%%%%%%%%%%%%%%%%  Figure 3  %%%%%%%%%%%%%%%

The calibration of the neural network is computationally expensive. With two input 
variables ($Z$ and $N$) and a ``canonical'' number of $H\!=\!40$ hidden nodes, a 
total of $1\!+\!4H\!=\!161$ parameters must be calibrated. To do so, we rely on the
\emph{Flexible Bayesian Modeling} package by Neal described in detail in 
Ref.\,\cite{Neal1996}. After an initial thermalization phase consisting of 500 Monte-Carlo 
steps, a sampling set of 100 configurations is accumulated to determine statistical 
averages and their associated uncertainties. As a measure of the accuracy of our 
predictions, we report root-mean-square (rms) deviations relative to experiment; 
for simplicity, the comparison is done using exclusively average values. Due to the 
relative simplicity of the DZ10 model, we were able to implement the BNN refinement 
using two different schemes. The first scheme consists of a unique set of masses 
obtained from the average values generated by the new DZ10 calibration---without 
accounting for the uncertainties in the bare model parameters. The second scheme 
remedies such deficiency by effectively incorporating the distribution of DZ10 model 
parameters depicted in Fig.\,\ref{Fig2}. Such improved version is useful in assessing 
whether a recalibration of the bare model parameters may be necessary in other cases.
Fortunately, our results suggest that, at least in the case of DZ10, this is not required. 
Our results indicate that whereas there is indeed a dramatic improvement in the 
predictions of the bare DZ10 model after the BNN refinement---from 
$\sigma_{\rm rms}\!=\!0.552\,{\rm MeV}$ to 
$\sigma_{\rm rms}\!=\!0.292\,{\rm MeV}$---no significant changes are observed 
when one uses the proper statistical distribution of bare DZ10 parameters; for
for this latter case one obtains $\sigma_{\rm rms}\!=\!0.296\,{\rm MeV}$. 

As alluded earlier, for each of the 100 Monte-Carlo configurations obtained 
from the BNN refinement one computes at every MC step the mass of each 
individual nucleus. Having generated all nuclear masses in such a way, one 
may then compute the mass differences required to generate one- and 
two-nucleon separation energies. At each MC step one proceeds in this 
same exact fashion, until by the end of the 100 Monte-Carlo configurations 
one can finally extract average values and associated theoretical uncertainties 
for each nuclear observable. The outcome of such a procedure is displayed in 
Fig.\,\ref{Fig3} for neutron and proton drip lines, identified at the point in which 
the two-neutron and two-proton separation energies become negative. For the 
case of the bare DZ10 model, the drip lines are displayed by a solid (red) line. 
In contrast, BNN-improved DZ10 predictions produce \emph{drip bands}, as all 
the predictions are now accompanied by statistical uncertainties. Note that 
because of the Coulomb repulsion, the proton drip line is much closer to the 
valley of stability than the corresponding neutron drip line. Indeed, whereas 
the proton drip line has been experimentally established for a large number 
of nuclei (up to atomic number $Z\!=\!91$), the neutron drip line is only known 
up to $Z\!=\!8$, with ${}^{24}$O being the heaviest known oxygen isotope that 
remains stable against particle decay\,\cite{Thoennessen:2004}. Moreover, 
Fig.\,\ref{Fig3} displays the characteristic signature of shell closures at neutron 
magic numbers 50, 82, 126, and (predicts) 184. Finally, the figure encapsulates the 
enormous experimental challenges faced in mapping the neutron drip line. 
However, even if reaching the neutron drip line for heavy nuclei may not be 
feasible in the foreseeable future, it is essential to continue the experimental
quest to properly inform theoretical models. In turn, BNN approaches as the 
one advocated here may guide experimental searches for those ``few" critical 
nuclei that may reduce the large systematic uncertainty that currently plagues
theoretical models. 

%%%%%%%%%%%%%%%%  Figure 4  %%%%%%%%%%%%%%%
\begin{figure}[ht]
\vspace{-0.05in}
\includegraphics[width=1\columnwidth,height=8cm]{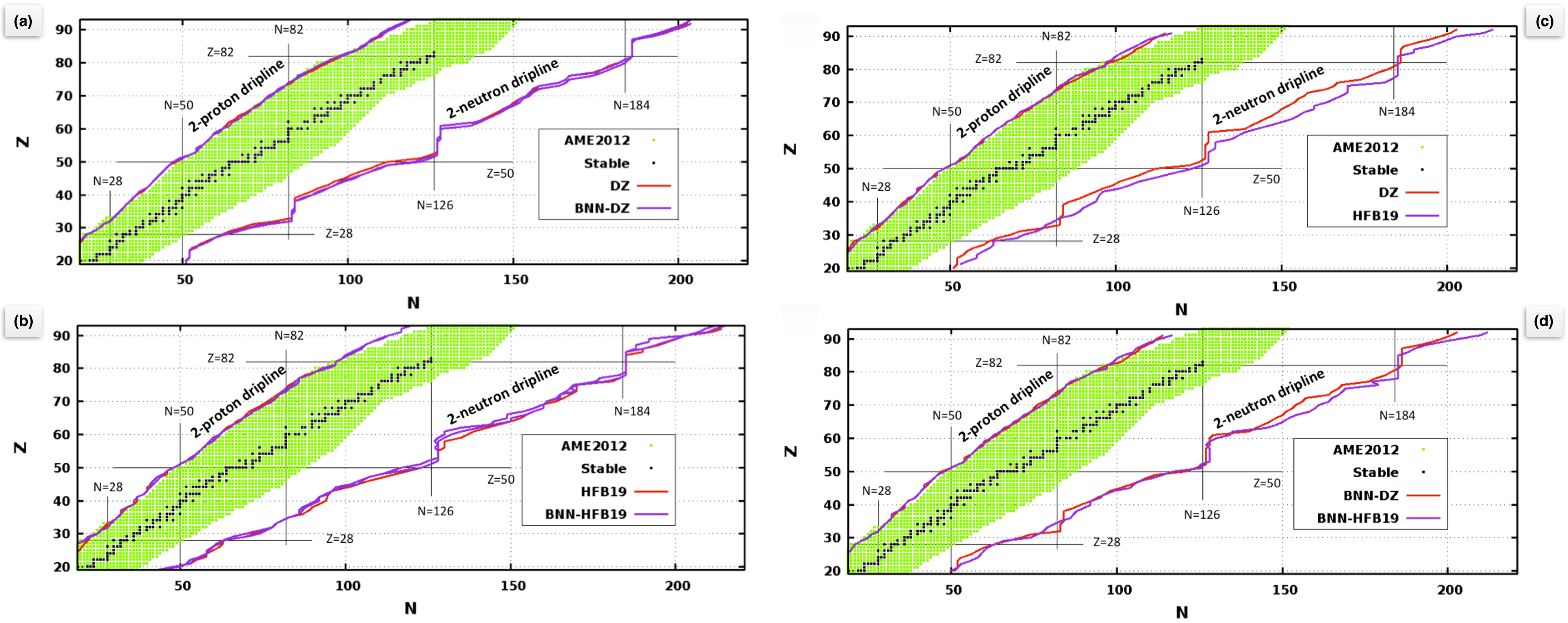} 
\caption{Proton and neutron drip lines/drip bands as predicted 
by (a) the Duflo-Zuker model\,\cite{Duflo:1995} and (b) the HFB19 
model\,\cite{Goriely:2010bm}; in both cases bare and BNN-refined 
predictions are displayed. The remaining two panels display the 
same information but now the comparison is between the 
predictions of both mass models: (c) displays bare-model results
whereas (d) the corresponding ones after BNN-refinement. Also 
shown (with green points) is the AME2012 data set\,\cite{AME:2012}
with the stable nuclei displayed with black points.}
\label{Fig4}
\end{figure}
%%%%%%%%%%%%%%%%  Figure 4  %%%%%%%%%%%%%%%

So far DZ10 has provided a benchmark to quantify the impact of the BNN
refinement and the role of uncertainty quantification. We found that while 
the BNN refinement is critical in improving the predictions of the model, there 
seems to be little value in propagating the uncertainties inherent to the bare 
model. Thus, we now proceed to document and test the predictions of the 
two state-of-the-art  mass models that will be refined in order to generate 
BNN-improved tables; the 28-parameter ``mic-mac'' model of Duflo 
and Zuker\,\cite{Duflo:1995} and the microscopic HFB19 model of Goriely,
Chamel, and Pearson\,\cite{Goriely:2010bm}. Drip lines and drip bands as
predicted by these two models are displayed in Fig.\,\ref{Fig4}. The two 
left-hand panels, (a) for Duflo-Zuker and (b) for HFB19, aim to illustrate 
the improvement to the bare models as a result of the BNN refinement.  
Note that although the overall improvement to both mass models is 
considerable---about 40\%\,\cite{Utama:2015hva}---the changes are difficult 
to discern because both the expanded scale as well as the intrinsic 
high quality of the bare models.  In contrast, the two 
right-hand panels compare the models to each other: (c) for the bare 
predictions and (d) for the BNN-improved versions. In this case the
improvement is clearly discernible, as the model discrepancies displayed
by the bare models get significantly quenched after the BNN refinement. 
Indeed, at first there is an appreciable discrepancy in the predictions of the 
bare models---with HFB19 consistently predicting the location of the neutron
drip line at larger values of $N$\!. Remarkably, much of the discrepancy 
disappears after the BNN refinement, especially in the region between 
shell closures $N=82$ and $N=126$. Such reduction in the systematic 
model error is highly desirable and entirely consistent with the results 
obtained in an earlier publication\,\cite{Utama:2015hva}. Moreover, we
observe a significant quantitative improvement in the predictions of both 
models. In the case of the Duflo-Zuker mass formula the root-mean-square 
mass deviation improves from $\sigma_{\rm rms}\!=\!0.503$\,MeV to 
$\sigma_{\rm rms}\!=\!0.286$\,MeV, whereas in the case of HFB19 it goes 
from $\sigma_{\rm rms}\!=\!0.559$\,MeV to $\sigma_{\rm rms}\!=\!0.358$\,MeV.

We close this section with a brief discussion of the impact of the BNN
refinement on a few critical $r$-process nuclei that emerge from the 
sensitivity study of Mumpower and collaborators\,\cite{Mumpower:2015ova,
Mumpower:2015zha,Mumpower:2015hva}; specifically, palladium ($Z\!=\!46$), 
cadmium ($Z\!=\!48$), indium ($Z\!=\!49$), and tin ($Z\!=\!50$). Note that these 
results may be readily extended to other isotopes by employing the refined mass 
tables provided here as supplemental material. The main goal of our analysis is 
to assess whether the ubiquitous bare-model discrepancies can be systematically 
reduced after the implementation of the BNN refinement. 

%%%%%%%%%%%%%%%%  Figure 5  %%%%%%%%%%%%%%%
\begin{figure}[h]
\vspace{-0.05in}
\includegraphics[width=0.8\columnwidth]{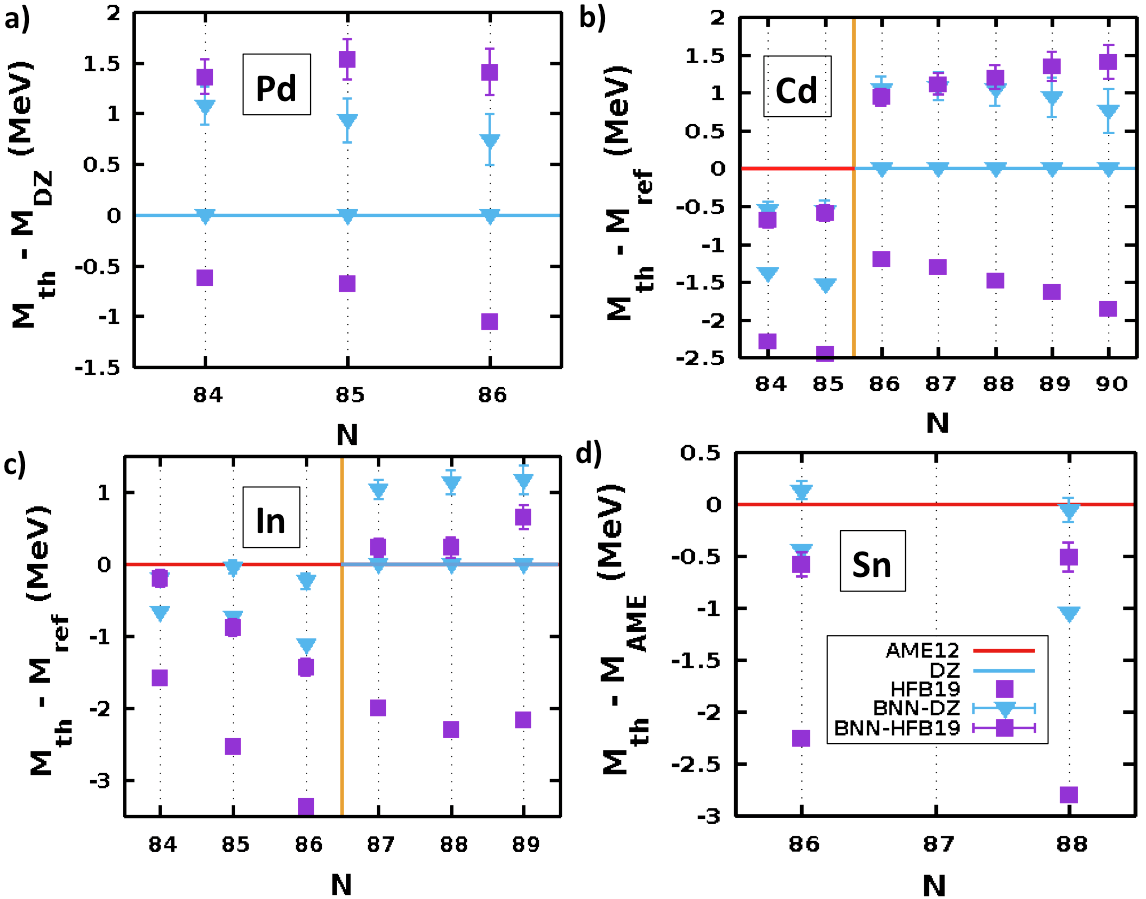}
\caption{Mass predictions for (a) palladium ($Z\!=\!46$), (b) cadmium ($Z\!=\!48$),
(c) indium ($Z\!=\!49$), and (d) tin ($Z\!=\!50$) for both the Duflo-Zuker\,\cite{Duflo:1995} 
and HFB19\,\cite{Goriely:2010bm} mass models. The predictions are relative to a reference 
mass value taken from either the AME2012 compilation when available\,\cite{AME:2012} or 
from the bare Duflo-Zuker model when unavailable. Predictions are displayed with statistical 
error bars for the BNN-improved models.}
\label{Fig5}
\end{figure}
%%%%%%%%%%%%%%%%  Figure 5  %%%%%%%%%%%%%%%

In Fig.\,\ref{Fig5} we display model predictions relative to a reference mass value 
for palladium, cadmium, indium, and tin in the region $N\!\gtrsim\!82$. The 
masses of all 18 nuclei displayed in the figure, ${}^{130-132}$Pd,
${}^{132-138}$Cd, ${}^{133-138}$In, and ${}^{136,138}$Sn, have been identified 
as important nuclei in the determination of the $r$-process abundance pattern; see 
Table I in Ref.\,\cite{Mumpower:2015hva}. Note that the reference mass has been 
adopted from either ``experimental" values (\emph{although derived not from purely 
experimental data}\,\cite{AME:2012}) or when data is unavailable, from the 
predictions of the \emph{bare} Duflo-Zuker model. Theoretical predictions with 
error bars are from the BNN-refined models. We can state categorically that 
the BNN refinement has an appreciable impact in reducing the systematic model 
discrepancies. Easiest to visualize are the cases of palladium and tin where 
the number of isotopes displayed is small; three and two, respectively. For 
palladium there is no available experimental data so we display theoretical 
estimates relative to the predictions from the bare Duflo-Zuker model (depicted 
by the light-blue line). Whereas the bare HFB19 predictions hover around 0.5-1 
MeV relative to such a baseline, the model discrepancy narrows considerably 
after the BNN refinement. Indeed, for ${}^{130}$Pd the predictions are now 
consistent with each other and for the most unfavorable case of ${}^{130}$Pd 
they agree at the 2$\sigma$ level. For tin the situation is similar, although in this 
case we display with a red line the recommended experimental 
values\,\cite{AME:2012}. Once again, we note that the bare-model predictions 
differ significantly from each other and, especially in the case of HFB19, from 
the recommended AME2012 values. However, once the BNN refinement is 
completed, both the model discrepancy is significantly reduced and the 
comparison with experiment is much more favorable. Indeed, both BNN-DZ 
predictions are now fully consistent with experiment. Finally, the two remaining 
isotopes of cadmium and indium display the same trends observed so far. In
particular, note that for all of these 18 important nuclei the BNN refinement 
suggest an \emph{increase} in the value of the mass, or equivalently, a 
reduction in the binding energy. For these two larger isotopic chains---having 
seven and six important nuclei, respectively---experimental mass values do 
exist but only for the smaller values of $N$. Yet regardless of whether 
experimental masses are available, we see the model spread diminishes 
considerably after the BNN refinement. And when recommended mass values 
exist, the BNN predictions are practically consistent with experiment. The isotopic 
chain in cadmium contains the largest number of important 
$r$-process nuclei. As we have learned from earlier studies (see for example 
Refs.\,\cite{Blaum:2006,Erler:2012,Mumpower:2015zha}) bare-model discrepancies 
diverge dramatically as one moves away from measured mass values. 
This is evident in Fig.\,\ref{Fig5}b where the bare-model discrepancy becomes as
large as $\sim\!2$\,MeV for ${}^{138}$Cd ({\sl i.e.,} $N=90$). Remarkably, all model 
discrepancies are largely eliminated after the BNN refinement. In particular, even 
for the least favorable case of ${}^{138}$Cd the 1$\sigma$ mismatch gets reduced 
to merely $130$\,keV. 

%%%%%%%%%%%%%%%%  Figure 6  %%%%%%%%%%%%%%%
\begin{figure}[h]
\vspace{-0.05in}
\includegraphics[width=0.8\columnwidth]{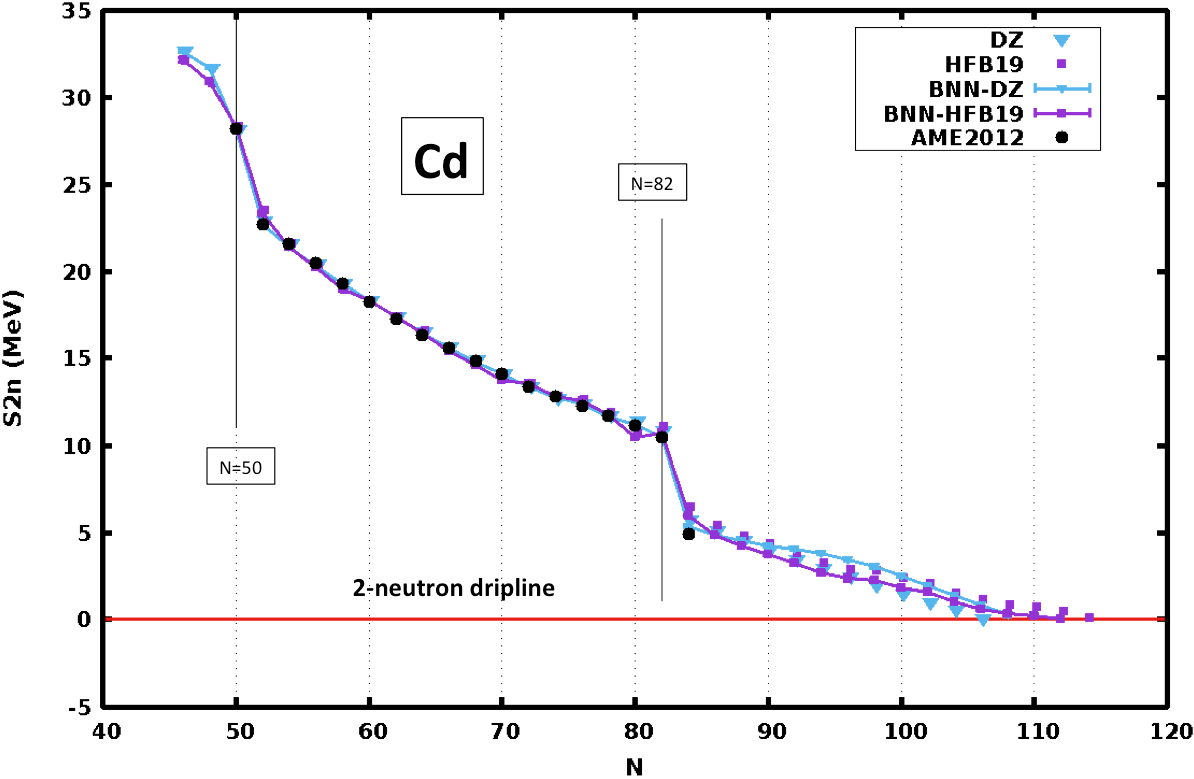}
\caption{Two-neutron separation energies for all even-even cadmium isotopes 
from $^{94}$Cd to $^{164}$Cd, as predicted by the mic-mac model of Duflo
and Zuker\,\cite{Duflo:1995} and the microscopic HFB19 model\,\cite{Goriely:2010bm}. 
Predictions are displayed without error bars for the bare models and with error bars 
after the BNN refinement. Experimental data are from Ref.\,\cite{AME:2012}.}
\label{Fig6}
\end{figure}
%%%%%%%%%%%%%%%%  Figure 6  %%%%%%%%%%%%%%%

We conclude this section by displaying in Fig.\,\ref{Fig6} two-neutron separation 
energies for all even-even isotopes in cadmium: from $^{94}$Cd to $^{164}$Cd. 
Clearly discernible in the figure is the characteristic ``jumps" at magic numbers 
50 and 82. The figure also encapsulates the inherent risk in extrapolating theoretical 
models far away from regions where experimental information is available. Whereas 
the models agree where data is available (from $50\!\le\!N\!\le\!82$), the 
model-discrepancy grows gradually with increasing neutron numbers. And while 
the BNN refinement is successful in mitigating the problem up to $N\!\lesssim\!90$ 
(see Fig.\,\ref{Fig5}c), the BNN predictions for $S_{2n}$ can differ by 
$\sim\!1$\,MeV far away from stability. This situation is reminiscent of the one
addressed in Ref.\,\cite{Erler:2012} for $S_{2n}$ in the case of the even-even 
erbium ($Z\!=\!68$) isotopes. It was found there that the discrepancy between various 
model predictions steadily grows with neutron excess as a result of the poorly 
known isovector effective interaction. Indeed, models that are successful in 
reproducing the experimentally determined two-neutron separation energy, differ 
in their predictions of the location of  the neutron drip line by as many as eight 
neutrons. Fortunately, next-generation rare-isotope facilities will produce hundreds 
of new exotic nuclei very far away from stability that will help constrain the isovector
sector. Yet a most pressing challenge is to identify the ``few" critical nuclei that 
will best inform nuclear theory. We are confident that the BNN formalism will 
successfully meet this challenge.

%%%%%%%%%%%%%%%%%%%%%%%%%%%%%%%%%%%%%%%%%%%%%%%%%%%%%%%%%%%%%%%%%%%%%%%%%%

\section{Conclusions}
\label{Conclusions}

The quest for the nuclear driplines is at the heart of one of the central themes 
animating nuclear physics today: \emph{what are the limits of nuclear existence?} 
As a fundamental nuclear-structure problem, the driplines define the most extreme
combinations of protons and neutrons that can remain bound by the strong nuclear 
force. Drip-line nuclei are weakly bound, finite, strongly correlated quantum systems 
where the virtual coupling to the continuum is critical. Besides being of intrinsic 
interest in nuclear structure, nuclei far away from stability also play a predominant role 
in astrophysics; for example, in stellar nucleosynthesis and in neutron stars. However, 
the enormous theoretical challenges faced in describing exotic nuclei are compounded 
by the lack of experimental guidance. For example, theoretical predictions seem to 
suggest that ${}^{118}$Kr is the drip-line nucleus separating the inner and outer crust 
of neutron stars. Yet ${}^{118}$Kr is 21 neutrons away from the last well-measured 
isotope. Thus, having to resort to extrapolations seems unavoidable.

Undoubtedly, large extrapolations pose a considerable risk. In an effort to mitigate
such risk we proposed a novel approach based on the construction of an artificial
neural network. Moreover, given that the neural-network parameters are determined
through Bayesian inference, our results are accompanied by statistical uncertainties.
The proposed approach adopts a highly successful mass model that is then refined 
through the construction of an artificial neural network. Briefly, the implementation of 
the BNN refinement proceeds by dividing the existing experimental database of nuclear 
masses into a learning and a validation set, with the members of each set selected at 
random. One then uses the learning set to train the artificial neural network by focusing 
on the mass residuals, {\sl i.e.,} on the difference between the experimentally measured 
and predicted mass. The validation set is then used to test the robustness and reliability 
of the refined model. If a significant improvement relative to the bare model is observed, 
then the training is repeated but now using the entire experimental database. Although 
we developed most of these ideas using the relatively simple, yet highly accurate, 
10-parameter Duflo-Zuker model, our ultimate goal was to publish refined mass tables for
two existing mass models, one microscopic and the other one of the mic-mac type. For the 
latter we used the 28-parameter Duflo-Zuker model while for the former HFB19. 

These are some of the most salient conclusions of our work. First, despite the intrinsic high-quality 
of both mass models, the BNN refinement lead to a significant improvement: 
$\sigma_{\rm rms}\!=\!(0.503\rightarrow0.286)$\,MeV and
$\sigma_{\rm rms}\!=\!(0.559\rightarrow0.358)$\,MeV for DZ and HFB19, respectively.
Second, although theoretical mass models agree in regions where masses are known
but differ widely in regions where they are not, we found a systematic and significant 
reduction in the model spread after implementing the BNN refinement. Third, given  the 
\emph{Bayesian} character of the approach, all BNN refined predictions are now accompanied 
by theoretical uncertainties. Finally, we provided as supplemental material mass tables, as well as 
tables for other derived quantities such as separation energies, for the two BNN refined models
considered in this work. We trust that these refined mass tables will be helpful for future 
astrophysical applications.

%%%%%%%%%%%%%%%%%%%%%%%%%%%%%%%%%%%%%%%%%%%%%%%%%%%%%%%%%%%%%%%%%%%%%%%%%%%%%

\begin{acknowledgments}
 This material is based upon work supported by the U.S. Department of Energy Office of Science, 
 Office of Nuclear Physics under Award Number DE-FD05-92ER40750.
\end{acknowledgments}
\vfill\eject

%%%%%%%%%%%%%%%%%%%%%%%%%%%%%%%%%%%%%%%%%%%%%%%%%%%%%%%%%%%%%%%%%%%%%%%%%%%%%%

\bibliography{../ReferencesJP.bib}

%merlin.mbs apsrev4-1.bst 2010-07-25 4.21a (PWD, AO, DPC) hacked
%Control: key (0)
%Control: author (8) initials jnrlst
%Control: editor formatted (1) identically to author
%Control: production of article title (-1) disabled
%Control: page (0) single
%Control: year (1) truncated
%Control: production of eprint (0) enabled
\begin{thebibliography}{55}%
\makeatletter
\providecommand \@ifxundefined [1]{%
 \@ifx{#1\undefined}
}%
\providecommand \@ifnum [1]{%
 \ifnum #1\expandafter \@firstoftwo
 \else \expandafter \@secondoftwo
 \fi
}%
\providecommand \@ifx [1]{%
 \ifx #1\expandafter \@firstoftwo
 \else \expandafter \@secondoftwo
 \fi
}%
\providecommand \natexlab [1]{#1}%
\providecommand \enquote  [1]{``#1''}%
\providecommand \bibnamefont  [1]{#1}%
\providecommand \bibfnamefont [1]{#1}%
\providecommand \citenamefont [1]{#1}%
\providecommand \href@noop [0]{\@secondoftwo}%
\providecommand \href [0]{\begingroup \@sanitize@url \@href}%
\providecommand \@href[1]{\@@startlink{#1}\@@href}%
\providecommand \@@href[1]{\endgroup#1\@@endlink}%
\providecommand \@sanitize@url [0]{\catcode `\\12\catcode `\$12\catcode
  `\&12\catcode `\#12\catcode `\^12\catcode `\_12\catcode `\%12\relax}%
\providecommand \@@startlink[1]{}%
\providecommand \@@endlink[0]{}%
\providecommand \url  [0]{\begingroup\@sanitize@url \@url }%
\providecommand \@url [1]{\endgroup\@href {#1}{\urlprefix }}%
\providecommand \urlprefix  [0]{URL }%
\providecommand \Eprint [0]{\href }%
\providecommand \doibase [0]{http://dx.doi.org/}%
\providecommand \selectlanguage [0]{\@gobble}%
\providecommand \bibinfo  [0]{\@secondoftwo}%
\providecommand \bibfield  [0]{\@secondoftwo}%
\providecommand \translation [1]{[#1]}%
\providecommand \BibitemOpen [0]{}%
\providecommand \bibitemStop [0]{}%
\providecommand \bibitemNoStop [0]{.\EOS\space}%
\providecommand \EOS [0]{\spacefactor3000\relax}%
\providecommand \BibitemShut  [1]{\csname bibitem#1\endcsname}%
\let\auto@bib@innerbib\@empty
%</preamble>
\bibitem [{Lon(2015)}]{LongRangePlan}%
  \BibitemOpen
  \href@noop {} {\emph {\bibinfo {title} {Reaching for the Horizon; The 2015
  Long Range Plan for Nuclear Science}}} (\bibinfo {year} {2015})\BibitemShut
  {NoStop}%
\bibitem [{\citenamefont {Clayton}(1983)}]{Clayton:1983}%
  \BibitemOpen
  \bibfield  {author} {\bibinfo {author} {\bibfnamefont {D.~D.}\ \bibnamefont
  {Clayton}},\ }\enquote {\bibinfo {title} {Principles of stellar evolution and
  nucleosynthesis},}\ \ (\bibinfo  {publisher} {University of Chicago Press},\
  \bibinfo {address} {Chicago},\ \bibinfo {year} {1983})\BibitemShut {NoStop}%
\bibitem [{\citenamefont {Phillips}(1998)}]{Phillips1998}%
  \BibitemOpen
  \bibfield  {author} {\bibinfo {author} {\bibfnamefont {A.~C.}\ \bibnamefont
  {Phillips}},\ }\enquote {\bibinfo {title} {The physics of stars},}\ \
  (\bibinfo  {publisher} {John Wiley \& Sons, Chichester},\ \bibinfo {year}
  {1998})\ \bibinfo {edition} {2nd}\ ed.\BibitemShut {Stop}%
\bibitem [{\citenamefont {Iliadis}(2007)}]{Iliadis:2007}%
  \BibitemOpen
  \bibfield  {author} {\bibinfo {author} {\bibfnamefont {C.}~\bibnamefont
  {Iliadis}},\ }\enquote {\bibinfo {title} {Nuclear physics of stars},}\ \
  (\bibinfo  {publisher} {Wiley-VCH},\ \bibinfo {address} {Weinheim},\ \bibinfo
  {year} {2007})\BibitemShut {NoStop}%
\bibitem [{\citenamefont {Burbidge}\ \emph {et~al.}(1957)\citenamefont
  {Burbidge}, \citenamefont {Burbidge}, \citenamefont {Fowler},\ and\
  \citenamefont {Hoyle}}]{Burbidge:1957vc}%
  \BibitemOpen
  \bibfield  {author} {\bibinfo {author} {\bibfnamefont {M.~E.}\ \bibnamefont
  {Burbidge}}, \bibinfo {author} {\bibfnamefont {G.~R.}\ \bibnamefont
  {Burbidge}}, \bibinfo {author} {\bibfnamefont {W.~A.}\ \bibnamefont
  {Fowler}}, \ and\ \bibinfo {author} {\bibfnamefont {F.}~\bibnamefont
  {Hoyle}},\ }\href {\doibase 10.1103/RevModPhys.29.547} {\bibfield  {journal}
  {\bibinfo  {journal} {Rev. Mod. Phys.}\ }\textbf {\bibinfo {volume} {29}},\
  \bibinfo {pages} {547} (\bibinfo {year} {1957})}\BibitemShut {NoStop}%
%%CITATION = RMPHA,29,547;%%
\bibitem [{\citenamefont {Wallerstein}\ \emph {et~al.}(1997)\citenamefont
  {Wallerstein}, \citenamefont {Iben}, \citenamefont {Parker}, \citenamefont
  {Boesgaard}, \citenamefont {Hale}, \citenamefont {Champagne} \emph
  {et~al.}}]{Wallerstein:1997}%
  \BibitemOpen
  \bibfield  {author} {\bibinfo {author} {\bibfnamefont {G.}~\bibnamefont
  {Wallerstein}}, \bibinfo {author} {\bibfnamefont {I.}~\bibnamefont {Iben}},
  \bibinfo {author} {\bibfnamefont {P.}~\bibnamefont {Parker}}, \bibinfo
  {author} {\bibfnamefont {A.~M.}\ \bibnamefont {Boesgaard}}, \bibinfo {author}
  {\bibfnamefont {G.~M.}\ \bibnamefont {Hale}}, \bibinfo {author}
  {\bibfnamefont {A.~E.}\ \bibnamefont {Champagne}},  \emph {et~al.},\ }\href
  {\doibase 10.1103/RevModPhys.69.995} {\bibfield  {journal} {\bibinfo
  {journal} {Rev. Mod. Phys.}\ }\textbf {\bibinfo {volume} {69}},\ \bibinfo
  {pages} {995} (\bibinfo {year} {1997})}\BibitemShut {NoStop}%
\bibitem [{Qua(2003)}]{QuarksCosmos:2003}%
  \BibitemOpen
  \href@noop {} {\emph {\bibinfo {title} {Connecting Quarks with the Cosmos:
  Eleven Science Questions for the New Century}}}\ (\bibinfo  {publisher} {The
  National Academies Press},\ \bibinfo {address} {Washington},\ \bibinfo {year}
  {2003})\BibitemShut {NoStop}%
\bibitem [{\citenamefont {Petermann}\ \emph {et~al.}(2010)\citenamefont
  {Petermann}, \citenamefont {Mart\'inez-Pinedo}, \citenamefont {Arcones},
  \citenamefont {Hix}, \citenamefont {Kelić}, \citenamefont {Langanke},
  \citenamefont {Panov}, \citenamefont {Rauscher}, \citenamefont {Schmidt},
  \citenamefont {Thielemann},\ and\ \citenamefont {Zinner}}]{Petermann:2010}%
  \BibitemOpen
  \bibfield  {author} {\bibinfo {author} {\bibfnamefont {I.}~\bibnamefont
  {Petermann}}, \bibinfo {author} {\bibfnamefont {G.}~\bibnamefont
  {Mart\'inez-Pinedo}}, \bibinfo {author} {\bibfnamefont {A.}~\bibnamefont
  {Arcones}}, \bibinfo {author} {\bibfnamefont {W.~R.}\ \bibnamefont {Hix}},
  \bibinfo {author} {\bibfnamefont {A.}~\bibnamefont {Kelić}}, \bibinfo
  {author} {\bibfnamefont {K.}~\bibnamefont {Langanke}}, \bibinfo {author}
  {\bibfnamefont {I.}~\bibnamefont {Panov}}, \bibinfo {author} {\bibfnamefont
  {T.}~\bibnamefont {Rauscher}}, \bibinfo {author} {\bibfnamefont {K.-H.}\
  \bibnamefont {Schmidt}}, \bibinfo {author} {\bibfnamefont {F.-K.}\
  \bibnamefont {Thielemann}}, \ and\ \bibinfo {author} {\bibfnamefont
  {N.}~\bibnamefont {Zinner}},\ }\href@noop {} {\bibfield  {journal} {\bibinfo
  {journal} {Journal of Physics: Conference Series}\ }\textbf {\bibinfo
  {volume} {202}},\ \bibinfo {pages} {012008} (\bibinfo {year}
  {2010})}\BibitemShut {NoStop}%
\bibitem [{\citenamefont {Mumpower}\ \emph {et~al.}(2016)\citenamefont
  {Mumpower}, \citenamefont {Surman}, \citenamefont {McLaughlin},\ and\
  \citenamefont {Aprahamian}}]{Mumpower:2015ova}%
  \BibitemOpen
  \bibfield  {author} {\bibinfo {author} {\bibfnamefont {M.~R.}\ \bibnamefont
  {Mumpower}}, \bibinfo {author} {\bibfnamefont {R.}~\bibnamefont {Surman}},
  \bibinfo {author} {\bibfnamefont {G.~C.}\ \bibnamefont {McLaughlin}}, \ and\
  \bibinfo {author} {\bibfnamefont {A.}~\bibnamefont {Aprahamian}},\
  }\href@noop {} {\bibfield  {journal} {\bibinfo  {journal} {Prog. Part. Nucl.
  Phys.}\ }\textbf {\bibinfo {volume} {86}},\ \bibinfo {pages} {86} (\bibinfo
  {year} {2016})}\BibitemShut {NoStop}%
%%CITATION = ARXIV:1508.07352;%%
\bibitem [{\citenamefont {Mumpower}\ \emph
  {et~al.}(2015{\natexlab{a}})\citenamefont {Mumpower}, \citenamefont {Surman},
  \citenamefont {Fang}, \citenamefont {Beard},\ and\ \citenamefont
  {Aprahamian}}]{Mumpower:2015zha}%
  \BibitemOpen
  \bibfield  {author} {\bibinfo {author} {\bibfnamefont {M.}~\bibnamefont
  {Mumpower}}, \bibinfo {author} {\bibfnamefont {R.}~\bibnamefont {Surman}},
  \bibinfo {author} {\bibfnamefont {D.~L.}\ \bibnamefont {Fang}}, \bibinfo
  {author} {\bibfnamefont {M.}~\bibnamefont {Beard}}, \ and\ \bibinfo {author}
  {\bibfnamefont {A.}~\bibnamefont {Aprahamian}},\ }\href {\doibase
  10.1088/0954-3899/42/3/034027} {\bibfield  {journal} {\bibinfo  {journal} {J.
  Phys.}\ }\textbf {\bibinfo {volume} {G42}},\ \bibinfo {pages} {034027}
  (\bibinfo {year} {2015}{\natexlab{a}})}\BibitemShut {NoStop}%
%%CITATION = JPAGA,G42,034027;%%
\bibitem [{\citenamefont {Mumpower}\ \emph
  {et~al.}(2015{\natexlab{b}})\citenamefont {Mumpower}, \citenamefont {Surman},
  \citenamefont {Fang}, \citenamefont {Beard}, \citenamefont {Moller},
  \citenamefont {Kawano},\ and\ \citenamefont {Aprahamian}}]{Mumpower:2015hva}%
  \BibitemOpen
  \bibfield  {author} {\bibinfo {author} {\bibfnamefont {M.~R.}\ \bibnamefont
  {Mumpower}}, \bibinfo {author} {\bibfnamefont {R.}~\bibnamefont {Surman}},
  \bibinfo {author} {\bibfnamefont {D.~L.}\ \bibnamefont {Fang}}, \bibinfo
  {author} {\bibfnamefont {M.}~\bibnamefont {Beard}}, \bibinfo {author}
  {\bibfnamefont {P.}~\bibnamefont {Moller}}, \bibinfo {author} {\bibfnamefont
  {T.}~\bibnamefont {Kawano}}, \ and\ \bibinfo {author} {\bibfnamefont
  {A.}~\bibnamefont {Aprahamian}},\ }\href {\doibase
  10.1103/PhysRevC.92.035807} {\bibfield  {journal} {\bibinfo  {journal} {Phys.
  Rev.}\ }\textbf {\bibinfo {volume} {C92}},\ \bibinfo {pages} {035807}
  (\bibinfo {year} {2015}{\natexlab{b}})}\BibitemShut {NoStop}%
%%CITATION = ARXIV:1505.07789;%%
\bibitem [{\citenamefont {Baym}\ \emph {et~al.}(1971)\citenamefont {Baym},
  \citenamefont {Pethick},\ and\ \citenamefont {Sutherland}}]{Baym:1971pw}%
  \BibitemOpen
  \bibfield  {author} {\bibinfo {author} {\bibfnamefont {G.}~\bibnamefont
  {Baym}}, \bibinfo {author} {\bibfnamefont {C.}~\bibnamefont {Pethick}}, \
  and\ \bibinfo {author} {\bibfnamefont {P.}~\bibnamefont {Sutherland}},\
  }\href@noop {} {\bibfield  {journal} {\bibinfo  {journal} {Astrophys. J.}\
  }\textbf {\bibinfo {volume} {170}},\ \bibinfo {pages} {299} (\bibinfo {year}
  {1971})}\BibitemShut {NoStop}%
%%CITATION = ASJOA,170,299;%%
\bibitem [{\citenamefont {Roca-Maza}\ and\ \citenamefont
  {Piekarewicz}(2008)}]{RocaMaza:2008ja}%
  \BibitemOpen
  \bibfield  {author} {\bibinfo {author} {\bibfnamefont {X.}~\bibnamefont
  {Roca-Maza}}\ and\ \bibinfo {author} {\bibfnamefont {J.}~\bibnamefont
  {Piekarewicz}},\ }\href@noop {} {\bibfield  {journal} {\bibinfo  {journal}
  {Phys. Rev.}\ }\textbf {\bibinfo {volume} {C78}},\ \bibinfo {pages} {025807}
  (\bibinfo {year} {2008})}\BibitemShut {NoStop}%
\bibitem [{\citenamefont {Utama}\ \emph
  {et~al.}(2016{\natexlab{a}})\citenamefont {Utama}, \citenamefont
  {Piekarewicz},\ and\ \citenamefont {Prosper}}]{Utama:2015hva}%
  \BibitemOpen
  \bibfield  {author} {\bibinfo {author} {\bibfnamefont {R.}~\bibnamefont
  {Utama}}, \bibinfo {author} {\bibfnamefont {J.}~\bibnamefont {Piekarewicz}},
  \ and\ \bibinfo {author} {\bibfnamefont {H.~B.}\ \bibnamefont {Prosper}},\
  }\href {\doibase 10.1103/PhysRevC.93.014311} {\bibfield  {journal} {\bibinfo
  {journal} {Phys. Rev.}\ }\textbf {\bibinfo {volume} {C93}},\ \bibinfo {pages}
  {014311} (\bibinfo {year} {2016}{\natexlab{a}})}\BibitemShut {NoStop}%
%%CITATION = ARXIV:1508.06263;%%
\bibitem [{\citenamefont {Piekarewicz}\ and\ \citenamefont
  {Utama}(2016)}]{Piekarewicz:2016akv}%
  \BibitemOpen
  \bibfield  {author} {\bibinfo {author} {\bibfnamefont {J.}~\bibnamefont
  {Piekarewicz}}\ and\ \bibinfo {author} {\bibfnamefont {R.}~\bibnamefont
  {Utama}},\ }\bibfield  {booktitle} {\emph {\bibinfo {booktitle}
  {{Proceedings, 34th Mazurian Lakes Conference on Physics: Frontiers in
  Nuclear Physics: Piaski, Poland, September 6-13, 2015}}},\ }\href {\doibase
  10.5506/APhysPolB.47.659} {\bibfield  {journal} {\bibinfo  {journal} {Acta
  Phys. Polon.}\ }\textbf {\bibinfo {volume} {B47}},\ \bibinfo {pages} {659}
  (\bibinfo {year} {2016})}\BibitemShut {NoStop}%
%%CITATION = APPOA,B47,659;%%
\bibitem [{\citenamefont {Utama}\ \emph
  {et~al.}(2016{\natexlab{b}})\citenamefont {Utama}, \citenamefont {Chen},\
  and\ \citenamefont {Piekarewicz}}]{Utama:2016rad}%
  \BibitemOpen
  \bibfield  {author} {\bibinfo {author} {\bibfnamefont {R.}~\bibnamefont
  {Utama}}, \bibinfo {author} {\bibfnamefont {W.-C.}\ \bibnamefont {Chen}}, \
  and\ \bibinfo {author} {\bibfnamefont {J.}~\bibnamefont {Piekarewicz}},\
  }\href@noop {} {\bibfield  {journal} {\bibinfo  {journal} {J. Phys.}\
  }\textbf {\bibinfo {volume} {G}} (\bibinfo {year}
  {2016}{\natexlab{b}})}\BibitemShut {NoStop}%
\bibitem [{\citenamefont {Blaum}(2006)}]{Blaum:2006}%
  \BibitemOpen
  \bibfield  {author} {\bibinfo {author} {\bibfnamefont {K.}~\bibnamefont
  {Blaum}},\ }\href@noop {} {\bibfield  {journal} {\bibinfo  {journal} {Physics
  Reports}\ }\textbf {\bibinfo {volume} {425}},\ \bibinfo {pages} {1} (\bibinfo
  {year} {2006})}\BibitemShut {NoStop}%
\bibitem [{\citenamefont {M\"oller}\ and\ \citenamefont
  {Nix}(1981)}]{Moller:1981zz}%
  \BibitemOpen
  \bibfield  {author} {\bibinfo {author} {\bibfnamefont {P.}~\bibnamefont
  {M\"oller}}\ and\ \bibinfo {author} {\bibfnamefont {J.~R.}\ \bibnamefont
  {Nix}},\ }\href@noop {} {\bibfield  {journal} {\bibinfo  {journal} {Atom.
  Data Nucl. Data Tabl.}\ }\textbf {\bibinfo {volume} {26}},\ \bibinfo {pages}
  {165} (\bibinfo {year} {1981})}\BibitemShut {NoStop}%
\bibitem [{\citenamefont {M\"oller}\ and\ \citenamefont
  {Nix}(1988)}]{Moller:1988}%
  \BibitemOpen
  \bibfield  {author} {\bibinfo {author} {\bibfnamefont {P.}~\bibnamefont
  {M\"oller}}\ and\ \bibinfo {author} {\bibfnamefont {J.~R.}\ \bibnamefont
  {Nix}},\ }\href@noop {} {\bibfield  {journal} {\bibinfo  {journal} {Atom.
  Data Nucl. Data Tabl.}\ }\textbf {\bibinfo {volume} {39}},\ \bibinfo {pages}
  {213} (\bibinfo {year} {1988})}\BibitemShut {NoStop}%
\bibitem [{\citenamefont {M\"oller}\ \emph {et~al.}(1995)\citenamefont
  {M\"oller}, \citenamefont {Nix}, \citenamefont {Myers},\ and\ \citenamefont
  {Swiatecki}}]{Moller:1993ed}%
  \BibitemOpen
  \bibfield  {author} {\bibinfo {author} {\bibfnamefont {P.}~\bibnamefont
  {M\"oller}}, \bibinfo {author} {\bibfnamefont {J.~R.}\ \bibnamefont {Nix}},
  \bibinfo {author} {\bibfnamefont {W.~D.}\ \bibnamefont {Myers}}, \ and\
  \bibinfo {author} {\bibfnamefont {W.~J.}\ \bibnamefont {Swiatecki}},\
  }\href@noop {} {\bibfield  {journal} {\bibinfo  {journal} {Atom. Data Nucl.
  Data Tabl.}\ }\textbf {\bibinfo {volume} {59}},\ \bibinfo {pages} {185}
  (\bibinfo {year} {1995})}\BibitemShut {NoStop}%
\bibitem [{\citenamefont {Duflo}\ and\ \citenamefont
  {Zuker}(1995)}]{Duflo:1995}%
  \BibitemOpen
  \bibfield  {author} {\bibinfo {author} {\bibfnamefont {J.}~\bibnamefont
  {Duflo}}\ and\ \bibinfo {author} {\bibfnamefont {A.}~\bibnamefont {Zuker}},\
  }\href {\doibase 10.1103/PhysRevC.52.R23} {\bibfield  {journal} {\bibinfo
  {journal} {Phys. Rev. C}\ }\textbf {\bibinfo {volume} {52}},\ \bibinfo
  {pages} {R23} (\bibinfo {year} {1995})}\BibitemShut {NoStop}%
\bibitem [{\citenamefont {Goriely}\ \emph {et~al.}(2010)\citenamefont
  {Goriely}, \citenamefont {Chamel},\ and\ \citenamefont
  {Pearson}}]{Goriely:2010bm}%
  \BibitemOpen
  \bibfield  {author} {\bibinfo {author} {\bibfnamefont {S.}~\bibnamefont
  {Goriely}}, \bibinfo {author} {\bibfnamefont {N.}~\bibnamefont {Chamel}}, \
  and\ \bibinfo {author} {\bibfnamefont {J.}~\bibnamefont {Pearson}},\ }\href
  {\doibase 10.1103/PhysRevC.82.035804} {\bibfield  {journal} {\bibinfo
  {journal} {Phys. Rev.}\ }\textbf {\bibinfo {volume} {C82}},\ \bibinfo {pages}
  {035804} (\bibinfo {year} {2010})}\BibitemShut {NoStop}%
\bibitem [{\citenamefont {Kortelainen}\ \emph {et~al.}()\citenamefont
  {Kortelainen}, \citenamefont {Lesinski}, \citenamefont {More}, \citenamefont
  {Nazarewicz}, \citenamefont {Sarich} \emph {et~al.}}]{Kortelainen:2010hv}%
  \BibitemOpen
  \bibfield  {author} {\bibinfo {author} {\bibfnamefont {M.}~\bibnamefont
  {Kortelainen}}, \bibinfo {author} {\bibfnamefont {T.}~\bibnamefont
  {Lesinski}}, \bibinfo {author} {\bibfnamefont {J.}~\bibnamefont {More}},
  \bibinfo {author} {\bibfnamefont {W.}~\bibnamefont {Nazarewicz}}, \bibinfo
  {author} {\bibfnamefont {J.}~\bibnamefont {Sarich}},  \emph {et~al.},\
  }\href@noop {} {\bibfield  {journal} {\bibinfo  {journal} {Phys.Rev.}\
  }\textbf {\bibinfo {volume} {C82}},\ \bibinfo {pages} {024313}}\BibitemShut
  {NoStop}%
\bibitem [{\citenamefont {Erler}\ \emph {et~al.}(2013)\citenamefont {Erler},
  \citenamefont {Horowitz}, \citenamefont {Nazarewicz}, \citenamefont
  {Rafalski},\ and\ \citenamefont {Reinhard}}]{Erler:2012qd}%
  \BibitemOpen
  \bibfield  {author} {\bibinfo {author} {\bibfnamefont {J.}~\bibnamefont
  {Erler}}, \bibinfo {author} {\bibfnamefont {C.~J.}\ \bibnamefont {Horowitz}},
  \bibinfo {author} {\bibfnamefont {W.}~\bibnamefont {Nazarewicz}}, \bibinfo
  {author} {\bibfnamefont {M.}~\bibnamefont {Rafalski}}, \ and\ \bibinfo
  {author} {\bibfnamefont {P.-G.}\ \bibnamefont {Reinhard}},\ }\href {\doibase
  10.1103/PhysRevC.87.044320} {\bibfield  {journal} {\bibinfo  {journal} {Phys.
  Rev.}\ }\textbf {\bibinfo {volume} {C87}},\ \bibinfo {pages} {044320}
  (\bibinfo {year} {2013})}\BibitemShut {NoStop}%
\bibitem [{\citenamefont {{Haensel}}\ \emph {et~al.}(1989)\citenamefont
  {{Haensel}}, \citenamefont {{Zdunik}},\ and\ \citenamefont
  {{Dobaczewski}}}]{Haensel:1989}%
  \BibitemOpen
  \bibfield  {author} {\bibinfo {author} {\bibfnamefont {P.}~\bibnamefont
  {{Haensel}}}, \bibinfo {author} {\bibfnamefont {J.~L.}\ \bibnamefont
  {{Zdunik}}}, \ and\ \bibinfo {author} {\bibfnamefont {J.}~\bibnamefont
  {{Dobaczewski}}},\ }\href@noop {} {\bibfield  {journal} {\bibinfo  {journal}
  {Astron. Astrophys.}\ }\textbf {\bibinfo {volume} {222}},\ \bibinfo {pages}
  {353} (\bibinfo {year} {1989})}\BibitemShut {NoStop}%
\bibitem [{\citenamefont {Haensel}\ and\ \citenamefont
  {Pichon}(1994)}]{Haensel:1993zw}%
  \BibitemOpen
  \bibfield  {author} {\bibinfo {author} {\bibfnamefont {P.}~\bibnamefont
  {Haensel}}\ and\ \bibinfo {author} {\bibfnamefont {B.}~\bibnamefont
  {Pichon}},\ }\href@noop {} {\bibfield  {journal} {\bibinfo  {journal}
  {Astron. Astrophys.}\ }\textbf {\bibinfo {volume} {283}},\ \bibinfo {pages}
  {313} (\bibinfo {year} {1994})}\BibitemShut {NoStop}%
\bibitem [{\citenamefont {Ruester}\ \emph {et~al.}(2006)\citenamefont
  {Ruester}, \citenamefont {Hempel},\ and\ \citenamefont
  {Schaffner-Bielich}}]{Ruester:2005fm}%
  \BibitemOpen
  \bibfield  {author} {\bibinfo {author} {\bibfnamefont {S.~B.}\ \bibnamefont
  {Ruester}}, \bibinfo {author} {\bibfnamefont {M.}~\bibnamefont {Hempel}}, \
  and\ \bibinfo {author} {\bibfnamefont {J.}~\bibnamefont
  {Schaffner-Bielich}},\ }\href@noop {} {\bibfield  {journal} {\bibinfo
  {journal} {Phys. Rev.}\ }\textbf {\bibinfo {volume} {C73}},\ \bibinfo {pages}
  {035804} (\bibinfo {year} {2006})}\BibitemShut {NoStop}%
\bibitem [{\citenamefont {Roca-Maza}\ \emph {et~al.}(2011)\citenamefont
  {Roca-Maza}, \citenamefont {Piekarewicz}, \citenamefont {Garcia-Galvez},\
  and\ \citenamefont {Centelles}}]{RocaMaza:2011pk}%
  \BibitemOpen
  \bibfield  {author} {\bibinfo {author} {\bibfnamefont {X.}~\bibnamefont
  {Roca-Maza}}, \bibinfo {author} {\bibfnamefont {J.}~\bibnamefont
  {Piekarewicz}}, \bibinfo {author} {\bibfnamefont {T.}~\bibnamefont
  {Garcia-Galvez}}, \ and\ \bibinfo {author} {\bibfnamefont {M.}~\bibnamefont
  {Centelles}},\ }in\ \href@noop {} {\emph {\bibinfo {booktitle} {Neutron Star
  Crust}}},\ \bibinfo {editor} {edited by\ \bibinfo {editor} {\bibfnamefont
  {C.}~\bibnamefont {Bertulani}}\ and\ \bibinfo {editor} {\bibfnamefont
  {J.}~\bibnamefont {Piekarewicz}}}\ (\bibinfo  {publisher} {Nova Publishers},\
  \bibinfo {address} {New York},\ \bibinfo {year} {2011})\BibitemShut {NoStop}%
\bibitem [{\citenamefont {Wolf}\ \emph {et~al.}(2013)\citenamefont {Wolf} \emph
  {et~al.}}]{Wolf:2013ge}%
  \BibitemOpen
  \bibfield  {author} {\bibinfo {author} {\bibfnamefont {R.}~\bibnamefont
  {Wolf}} \emph {et~al.},\ }\href {\doibase 10.1103/PhysRevLett.110.041101}
  {\bibfield  {journal} {\bibinfo  {journal} {Phys. Rev. Lett.}\ }\textbf
  {\bibinfo {volume} {110}},\ \bibinfo {pages} {041101} (\bibinfo {year}
  {2013})}\BibitemShut {NoStop}%
%%CITATION = PRLTA,110,041101;%%
\bibitem [{\citenamefont {Wang}\ \emph
  {et~al.}(2012{\natexlab{a}})\citenamefont {Wang}, \citenamefont {Audi},
  \citenamefont {Wapstra}, \citenamefont {Kondev}, \citenamefont {MacCormick},
  \citenamefont {Xu},\ and\ \citenamefont {Pfeiffer}}]{AME:2012}%
  \BibitemOpen
  \bibfield  {author} {\bibinfo {author} {\bibfnamefont {M.}~\bibnamefont
  {Wang}}, \bibinfo {author} {\bibfnamefont {G.}~\bibnamefont {Audi}}, \bibinfo
  {author} {\bibfnamefont {A.}~\bibnamefont {Wapstra}}, \bibinfo {author}
  {\bibfnamefont {F.}~\bibnamefont {Kondev}}, \bibinfo {author} {\bibfnamefont
  {M.}~\bibnamefont {MacCormick}}, \bibinfo {author} {\bibfnamefont
  {X.}~\bibnamefont {Xu}}, \ and\ \bibinfo {author} {\bibfnamefont
  {B.}~\bibnamefont {Pfeiffer}},\ }\href {\doibase 10.1088/1674-1137/36/12/003}
  {\bibfield  {journal} {\bibinfo  {journal} {Chinese Phys. C}\ }\textbf
  {\bibinfo {volume} {36}},\ \bibinfo {pages} {1603} (\bibinfo {year}
  {2012}{\natexlab{a}})}\BibitemShut {NoStop}%
\bibitem [{\citenamefont {Neal}(1996)}]{Neal1996}%
  \BibitemOpen
  \bibfield  {author} {\bibinfo {author} {\bibfnamefont {R.}~\bibnamefont
  {Neal}},\ }\href@noop {} {\emph {\bibinfo {title} {Bayesian Learning of
  Neural Network}}}\ (\bibinfo  {publisher} {Springer},\ \bibinfo {address}
  {New York},\ \bibinfo {year} {1996})\BibitemShut {NoStop}%
\bibitem [{\citenamefont {Cybenko}(1989)}]{Cybenko:1989}%
  \BibitemOpen
  \bibfield  {author} {\bibinfo {author} {\bibfnamefont {G.}~\bibnamefont
  {Cybenko}},\ }\href@noop {} {\bibfield  {journal} {\bibinfo  {journal} {Math.
  Control Signals Systems}\ }\textbf {\bibinfo {volume} {2}},\ \bibinfo {pages}
  {303} (\bibinfo {year} {1989})}\BibitemShut {NoStop}%
\bibitem [{\citenamefont {Hornik}\ \emph {et~al.}(1989)\citenamefont {Hornik},
  \citenamefont {Stinchcombe},\ and\ \citenamefont {White}}]{Hornik:1989}%
  \BibitemOpen
  \bibfield  {author} {\bibinfo {author} {\bibfnamefont {K.}~\bibnamefont
  {Hornik}}, \bibinfo {author} {\bibfnamefont {M.}~\bibnamefont {Stinchcombe}},
  \ and\ \bibinfo {author} {\bibfnamefont {H.}~\bibnamefont {White}},\
  }\href@noop {} {\bibfield  {journal} {\bibinfo  {journal} {Neural Networks}\
  }\textbf {\bibinfo {volume} {2}},\ \bibinfo {pages} {359 } (\bibinfo {year}
  {1989})}\BibitemShut {NoStop}%
\bibitem [{\citenamefont {Bishop}()}]{Bishop1995}%
  \BibitemOpen
  \bibfield  {author} {\bibinfo {author} {\bibfnamefont {C.}~\bibnamefont
  {Bishop}},\ }\href@noop {} {\emph {\bibinfo {title} {Neural Networks for
  Pattern Recognition}}}\ (\bibinfo  {publisher} {Oxford University Press},\
  \bibinfo {address} {Birmingham, UK})\BibitemShut {NoStop}%
\bibitem [{\citenamefont {Haykin}()}]{Haykin1999}%
  \BibitemOpen
  \bibfield  {author} {\bibinfo {author} {\bibfnamefont {S.}~\bibnamefont
  {Haykin}},\ }\href@noop {} {\emph {\bibinfo {title} {Neural Networks: A
  Comprehensive Foundation}}}\ (\bibinfo  {publisher} {Prentice Hall},\
  \bibinfo {address} {Upper Saddle River, NJ})\BibitemShut {NoStop}%
\bibitem [{\citenamefont {Vapnik}()}]{Vapnik1998}%
  \BibitemOpen
  \bibfield  {author} {\bibinfo {author} {\bibfnamefont {V.}~\bibnamefont
  {Vapnik}},\ }\href@noop {} {\emph {\bibinfo {title} {Statistical Learning
  Theory}}}\ (\bibinfo  {publisher} {Wiley-Interscience},\ \bibinfo {address}
  {New York, NY})\BibitemShut {NoStop}%
\bibitem [{\citenamefont {Gazula}\ \emph {et~al.}(1992)\citenamefont {Gazula},
  \citenamefont {Clark},\ and\ \citenamefont {Bohr}}]{Gazula:1992}%
  \BibitemOpen
  \bibfield  {author} {\bibinfo {author} {\bibfnamefont {S.}~\bibnamefont
  {Gazula}}, \bibinfo {author} {\bibfnamefont {J.}~\bibnamefont {Clark}}, \
  and\ \bibinfo {author} {\bibfnamefont {H.}~\bibnamefont {Bohr}},\ }\href
  {\doibase 10.1016/0375-9474(92)90191-L} {\bibfield  {journal} {\bibinfo
  {journal} {Nucl. Phys. A}\ }\textbf {\bibinfo {volume} {540}},\ \bibinfo
  {pages} {1} (\bibinfo {year} {1992})}\BibitemShut {NoStop}%
\bibitem [{\citenamefont {Gernoth}\ \emph {et~al.}(1993)\citenamefont
  {Gernoth}, \citenamefont {Clark}, \citenamefont {Prater},\ and\ \citenamefont
  {Bohr}}]{Gernoth:1993}%
  \BibitemOpen
  \bibfield  {author} {\bibinfo {author} {\bibfnamefont {K.}~\bibnamefont
  {Gernoth}}, \bibinfo {author} {\bibfnamefont {J.}~\bibnamefont {Clark}},
  \bibinfo {author} {\bibfnamefont {J.}~\bibnamefont {Prater}}, \ and\ \bibinfo
  {author} {\bibfnamefont {H.}~\bibnamefont {Bohr}},\ }\href {\doibase
  10.1016/0370-2693(93)90738-4} {\bibfield  {journal} {\bibinfo  {journal}
  {Phys. Lett. B}\ }\textbf {\bibinfo {volume} {300}},\ \bibinfo {pages} {1}
  (\bibinfo {year} {1993})}\BibitemShut {NoStop}%
\bibitem [{\citenamefont {Gernoth}\ and\ \citenamefont
  {Clark}(1995)}]{Gernoth:1995}%
  \BibitemOpen
  \bibfield  {author} {\bibinfo {author} {\bibfnamefont {K.}~\bibnamefont
  {Gernoth}}\ and\ \bibinfo {author} {\bibfnamefont {J.}~\bibnamefont
  {Clark}},\ }\href@noop {} {\bibfield  {journal} {\bibinfo  {journal} {Neural
  Networks}\ }\textbf {\bibinfo {volume} {8}},\ \bibinfo {pages} {291}
  (\bibinfo {year} {1995})}\BibitemShut {NoStop}%
\bibitem [{\citenamefont {Clark}\ \emph {et~al.}(1999)\citenamefont {Clark},
  \citenamefont {Lindenau},\ and\ \citenamefont {Ristig}}]{Clark:1999}%
  \BibitemOpen
  \bibfield  {author} {\bibinfo {author} {\bibfnamefont {J.~W.}\ \bibnamefont
  {Clark}}, \bibinfo {author} {\bibfnamefont {T.}~\bibnamefont {Lindenau}}, \
  and\ \bibinfo {author} {\bibfnamefont {M.}~\bibnamefont {Ristig}},\ }\enquote
  {\bibinfo {title} {Scientific applications of neural nets springer lecture
  notes in physics},}\ \ (\bibinfo  {publisher} {Springer-Verlag},\ \bibinfo
  {address} {Berlin},\ \bibinfo {year} {1999})\ pp.\ \bibinfo {pages}
  {1--96}\BibitemShut {NoStop}%
\bibitem [{\citenamefont {Athanassopoulos}\ \emph {et~al.}(2004)\citenamefont
  {Athanassopoulos}, \citenamefont {Mavrommatis}, \citenamefont {Gernoth},\
  and\ \citenamefont {Clark}}]{Athanassopoulos:2003qe}%
  \BibitemOpen
  \bibfield  {author} {\bibinfo {author} {\bibfnamefont {S.}~\bibnamefont
  {Athanassopoulos}}, \bibinfo {author} {\bibfnamefont {E.}~\bibnamefont
  {Mavrommatis}}, \bibinfo {author} {\bibfnamefont {K.~A.}\ \bibnamefont
  {Gernoth}}, \ and\ \bibinfo {author} {\bibfnamefont {J.~W.}\ \bibnamefont
  {Clark}},\ }\href@noop {} {\bibfield  {journal} {\bibinfo  {journal} {Nucl.
  Phys.}\ }\textbf {\bibinfo {volume} {A743}},\ \bibinfo {pages} {222}
  (\bibinfo {year} {2004})}\BibitemShut {NoStop}%
\bibitem [{\citenamefont {Athanassopoulos}\ \emph {et~al.}(2006)\citenamefont
  {Athanassopoulos}, \citenamefont {Mavrommatis}, \citenamefont {Gernoth},\
  and\ \citenamefont {Clark}}]{Athanassopoulos:2005rc}%
  \BibitemOpen
  \bibfield  {author} {\bibinfo {author} {\bibfnamefont {S.}~\bibnamefont
  {Athanassopoulos}}, \bibinfo {author} {\bibfnamefont {E.}~\bibnamefont
  {Mavrommatis}}, \bibinfo {author} {\bibfnamefont {K.~A.}\ \bibnamefont
  {Gernoth}}, \ and\ \bibinfo {author} {\bibfnamefont {J.~W.}\ \bibnamefont
  {Clark}},\ }\href@noop {} {\emph {\bibinfo {title} {Nuclear mass systematics
  by complementing the finite range droplet model with neural networks}}},\
  edited by\ \bibinfo {editor} {\bibfnamefont {G.}~\bibnamefont {Lalazissis}}\
  and\ \bibinfo {editor} {\bibfnamefont {C.}~\bibnamefont {Moustakidis}}\
  (\bibinfo  {publisher} {Advances in Nuclear Physics, Proceedings of the 15th
  Hellenic Symposium on Nuclear Physics},\ \bibinfo {year} {2006})\ pp.\
  \bibinfo {pages} {65--70}\BibitemShut {NoStop}%
\bibitem [{\citenamefont {Clark}\ and\ \citenamefont
  {Li}(2006)}]{Clark:2006ua}%
  \BibitemOpen
  \bibfield  {author} {\bibinfo {author} {\bibfnamefont {J.~W.}\ \bibnamefont
  {Clark}}\ and\ \bibinfo {author} {\bibfnamefont {H.}~\bibnamefont {Li}},\
  }\href@noop {} {\bibfield  {journal} {\bibinfo  {journal} {Int. J. Mod.
  Phys.}\ }\textbf {\bibinfo {volume} {B20}},\ \bibinfo {pages} {5015}
  (\bibinfo {year} {2006})}\BibitemShut {NoStop}%
%%CITATION = NUCL-TH/0603037;%%
\bibitem [{\citenamefont {Costiris}\ \emph {et~al.}(2009)\citenamefont
  {Costiris}, \citenamefont {Mavrommatis}, \citenamefont {Gernoth},\ and\
  \citenamefont {Clark}}]{Costiris:2009}%
  \BibitemOpen
  \bibfield  {author} {\bibinfo {author} {\bibfnamefont {N.~J.}\ \bibnamefont
  {Costiris}}, \bibinfo {author} {\bibfnamefont {E.}~\bibnamefont
  {Mavrommatis}}, \bibinfo {author} {\bibfnamefont {K.~A.}\ \bibnamefont
  {Gernoth}}, \ and\ \bibinfo {author} {\bibfnamefont {J.~W.}\ \bibnamefont
  {Clark}},\ }\href@noop {} {\bibfield  {journal} {\bibinfo  {journal} {Phys.
  Rev. C}\ }\textbf {\bibinfo {volume} {80}},\ \bibinfo {pages} {044332}
  (\bibinfo {year} {2009})}\BibitemShut {NoStop}%
\bibitem [{\citenamefont {Bayram}\ \emph {et~al.}(2014)\citenamefont {Bayram},
  \citenamefont {Akkoyun},\ and\ \citenamefont {Kara}}]{Bayram:2013hi}%
  \BibitemOpen
  \bibfield  {author} {\bibinfo {author} {\bibfnamefont {T.}~\bibnamefont
  {Bayram}}, \bibinfo {author} {\bibfnamefont {S.}~\bibnamefont {Akkoyun}}, \
  and\ \bibinfo {author} {\bibfnamefont {S.~O.}\ \bibnamefont {Kara}},\
  }\href@noop {} {\bibfield  {journal} {\bibinfo  {journal} {Annals of Nuclear
  Energy}\ }\textbf {\bibinfo {volume} {63}},\ \bibinfo {pages} {172} (\bibinfo
  {year} {2014})}\BibitemShut {NoStop}%
%%CITATION = ARXIV:1301.2407;%%
\bibitem [{\citenamefont {Stone}(2013)}]{Stone:2013}%
  \BibitemOpen
  \bibfield  {author} {\bibinfo {author} {\bibfnamefont {J.~V.}\ \bibnamefont
  {Stone}},\ }\enquote {\bibinfo {title} {Bayes' rule: A tutorial introduction
  to bayesian analysis},}\ \ (\bibinfo  {publisher} {Sebtel Press},\ \bibinfo
  {address} {Sheffield, UK},\ \bibinfo {year} {2013})\ \bibinfo {edition}
  {1st}\ ed.\BibitemShut {Stop}%
\bibitem [{\citenamefont {MacKay}(1995)}]{Mackay:1995}%
  \BibitemOpen
  \bibfield  {author} {\bibinfo {author} {\bibfnamefont {D.~J.}\ \bibnamefont
  {MacKay}},\ }\href@noop {} {\bibfield  {journal} {\bibinfo  {journal}
  {Nuclear Instruments and Methods in Physics Research Section A}\ }\textbf
  {\bibinfo {volume} {354}},\ \bibinfo {pages} {73 } (\bibinfo {year}
  {1995})}\BibitemShut {NoStop}%
\bibitem [{\citenamefont {MacKay}(1999)}]{Mackay:1999}%
  \BibitemOpen
  \bibfield  {author} {\bibinfo {author} {\bibfnamefont {D.~J.}\ \bibnamefont
  {MacKay}},\ }\href@noop {} {\bibfield  {journal} {\bibinfo  {journal} {Neural
  Computation}\ }\textbf {\bibinfo {volume} {11}},\ \bibinfo {pages} {1035}
  (\bibinfo {year} {1999})}\BibitemShut {NoStop}%
\bibitem [{\citenamefont {Mendoza-Temis}\ \emph {et~al.}(2010)\citenamefont
  {Mendoza-Temis}, \citenamefont {Hirsch},\ and\ \citenamefont
  {Zuker}}]{MendozaTemis:2009ia}%
  \BibitemOpen
  \bibfield  {author} {\bibinfo {author} {\bibfnamefont {J.}~\bibnamefont
  {Mendoza-Temis}}, \bibinfo {author} {\bibfnamefont {J.~G.}\ \bibnamefont
  {Hirsch}}, \ and\ \bibinfo {author} {\bibfnamefont {A.~P.}\ \bibnamefont
  {Zuker}},\ }\href@noop {} {\bibfield  {journal} {\bibinfo  {journal} {Nucl.
  Phys.}\ }\textbf {\bibinfo {volume} {A843}},\ \bibinfo {pages} {14} (\bibinfo
  {year} {2010})}\BibitemShut {NoStop}%
%%CITATION = ARXIV:0912.0882;%%
\bibitem [{\citenamefont {Audi}\ and\ \citenamefont
  {Wapstra}(1995)}]{Audi:1995}%
  \BibitemOpen
  \bibfield  {author} {\bibinfo {author} {\bibfnamefont {G.}~\bibnamefont
  {Audi}}\ and\ \bibinfo {author} {\bibfnamefont {A.~H.}\ \bibnamefont
  {Wapstra}},\ }\href@noop {} {\bibfield  {journal} {\bibinfo  {journal} {Nucl.
  Phys.}\ }\textbf {\bibinfo {volume} {A595}},\ \bibinfo {pages} {409}
  (\bibinfo {year} {1995})}\BibitemShut {NoStop}%
%%%%
\bibitem [{\citenamefont {Audi}\ \emph {et~al.}(2002)\citenamefont {Audi},
  \citenamefont {Wapstra},\ and\ \citenamefont {Thibault}}]{Audi:2002rp}%
  \BibitemOpen
  \bibfield  {author} {\bibinfo {author} {\bibfnamefont {G.}~\bibnamefont
  {Audi}}, \bibinfo {author} {\bibfnamefont {A.~H.}\ \bibnamefont {Wapstra}}, \
  and\ \bibinfo {author} {\bibfnamefont {C.}~\bibnamefont {Thibault}},\ }\href
  {\doibase 10.1016/j.nuclphysa.2003.11.003} {\bibfield  {journal} {\bibinfo
  {journal} {Nucl. Phys.}\ }\textbf {\bibinfo {volume} {A729}},\ \bibinfo
  {pages} {337} (\bibinfo {year} {2002})}\BibitemShut {NoStop}%
%%CITATION = NUPHA,A729,337;%%
\bibitem [{\citenamefont {Wang}\ \emph
  {et~al.}(2012{\natexlab{b}})\citenamefont {Wang}, \citenamefont {Audi},
  \citenamefont {Wapstra}, \citenamefont {Kondev}, \citenamefont {MacCormick},
  \citenamefont {Xu},\ and\ \citenamefont {Pfeiffer}}]{Wang:2012}%
  \BibitemOpen
  \bibfield  {author} {\bibinfo {author} {\bibfnamefont {M.}~\bibnamefont
  {Wang}}, \bibinfo {author} {\bibfnamefont {G.}~\bibnamefont {Audi}}, \bibinfo
  {author} {\bibfnamefont {A.~H.}\ \bibnamefont {Wapstra}}, \bibinfo {author}
  {\bibfnamefont {F.~G.}\ \bibnamefont {Kondev}}, \bibinfo {author}
  {\bibfnamefont {M.}~\bibnamefont {MacCormick}}, \bibinfo {author}
  {\bibfnamefont {X.}~\bibnamefont {Xu}}, \ and\ \bibinfo {author}
  {\bibfnamefont {B.}~\bibnamefont {Pfeiffer}},\ }\href@noop {} {\bibfield
  {journal} {\bibinfo  {journal} {Chinese Phys. C}\ }\textbf {\bibinfo {volume}
  {36}},\ \bibinfo {pages} {1603} (\bibinfo {year}
  {2012}{\natexlab{b}})}\BibitemShut {NoStop}%
\bibitem [{\citenamefont {Doboszewski}(2014)}]{Doboszewski:2014}%
  \BibitemOpen
  \bibfield  {author} {\bibinfo {author} {\bibfnamefont {I.}~\bibnamefont
  {Doboszewski}},\ }\emph {\bibinfo {title} {Predictive Power of Atomic Nuclei
  Mass Models}},\ \href@noop {} {Master's thesis},\ \bibinfo  {school} {AGH
  University of Science and Technology}, \bibinfo {address} {Cracow, Poland}
  (\bibinfo {year} {2014}),\ \bibinfo {note} {in Polish,
  unpublished}\BibitemShut {NoStop}%
\bibitem [{\citenamefont {Erler}\ \emph {et~al.}(2012)\citenamefont {Erler},
  \citenamefont {Birge}, \citenamefont {Kortelainen}, \citenamefont
  {Nazarewicz}, \citenamefont {Olsen}, \citenamefont {Perhac},\ and\
  \citenamefont {Stoitsov}}]{Erler:2012}%
  \BibitemOpen
  \bibfield  {author} {\bibinfo {author} {\bibfnamefont {J.}~\bibnamefont
  {Erler}}, \bibinfo {author} {\bibfnamefont {N.}~\bibnamefont {Birge}},
  \bibinfo {author} {\bibfnamefont {M.}~\bibnamefont {Kortelainen}}, \bibinfo
  {author} {\bibfnamefont {W.}~\bibnamefont {Nazarewicz}}, \bibinfo {author}
  {\bibfnamefont {E.}~\bibnamefont {Olsen}}, \bibinfo {author} {\bibfnamefont
  {A.~M.}\ \bibnamefont {Perhac}}, \ and\ \bibinfo {author} {\bibfnamefont
  {M.}~\bibnamefont {Stoitsov}},\ }\href@noop {} {\bibfield  {journal}
  {\bibinfo  {journal} {Nature}\ }\textbf {\bibinfo {volume} {486}},\ \bibinfo
  {pages} {509} (\bibinfo {year} {2012})}\BibitemShut {NoStop}%
\bibitem [{\citenamefont {Thoennessen}(2004)}]{Thoennessen:2004}%
  \BibitemOpen
  \bibfield  {author} {\bibinfo {author} {\bibfnamefont {M.}~\bibnamefont
  {Thoennessen}},\ }\href@noop {} {\bibfield  {journal} {\bibinfo  {journal}
  {Rep. Prog. Phys.}\ }\textbf {\bibinfo {volume} {67}},\ \bibinfo {pages}
  {1187} (\bibinfo {year} {2004})}\BibitemShut {NoStop}%
\end{thebibliography}%
%\bibliography{./BNNDriplines.bbl}

\end{document}